%% file: main.tex
\documentclass[twocolumn]{aastex631} % Use for publication
\usepackage[figure,figure*]{hypcap}
\usepackage{longtable,hyperref, graphicx, subfigure, comment}
\usepackage[fleqn]{amsmath}
\usepackage{amstext, float}
\usepackage{newtxmath} %use times font for math
\usepackage{lmodern}
\usepackage[T1]{fontenc}
\usepackage{url, footmisc}
\usepackage{soul}
\usepackage{booktabs}

\newcounter{column_number}
\newcommand{\appref}[1]{\hyperref[#1]{Appendix~\ref*{#1}}} % To get Appendix A.1 instead of Section A.1
\newcommand{\lapprox }{{\lower0.8ex\hbox{$\buildrel <\over\sim$}}}
\newcommand{\gapprox }{{\lower0.8ex\hbox{$\buildrel >\over\sim$}}}
\newcommand{\Msun}{\ifmmode {M_{\odot}}\else${M_{\odot}}$\fi}
\newcommand{\degree}{\ifmmode {^\circ}\else$^\circ$\fi}
\newcommand{\Ro}{\ifmmode {R_\mathrm{o}}\else$R_\mathrm{o}$\fi}

\def\gminusk{($G - K$)}
\def\amin{\ifmmode^{\prime}\else$^{\prime}$\fi}
\def\asec{\ifmmode^{\prime\prime}\else$^{\prime\prime}$\fi}
\def\halpha{$\mathrm{H}\alpha$}
\def\fx{$f_{\mathrm{X}}$}
\def\LX{$L_{\mathrm{X}}$}

\def\Lbol{$L_{\mathrm{bol}}$}
\def\LLX{$L_{\mathrm{X}}$/$L_{\mathrm{bol}}$}
\def\LLH{$L_{\mathrm{H}\alpha}$/$L_{\mathrm{bol}}$}
\newcommand{\prot}{\ensuremath{P_{\mbox{\scriptsize rot}}}}
\def\teff{$T_{\mathrm{eff}}$}

\defcitealias{Douglas2014}{Paper II}
\defcitealias{Nunez2022}{Paper IV}

\shorttitle{Chromospheric and Coronal Activity and Rotation in Praesepe and the Hyades}
\shortauthors{N\'u\~nez et al.}

\begin{document}

\title{\sc The Factory and the Beehive.~V.~Chromospheric and Coronal Activity and Its Dependence on Rotation in Praesepe and the Hyades}

\correspondingauthor{Alejandro N\'u{\~n}ez}
\email{alejo.nunez@gmail.com}

\newcommand{\columbia}{Department of Astronomy, Columbia University, 550 West 120th Street, New York, NY 10027, USA}

\newcommand{\LAB}{Laboratoire d'astrophysique de Bordeaux, Univ.~Bordeaux, CNRS, B18N, All\'ee Geoffroy Saint-Hilaire, 33615 Pessac, France}

\newcommand{\lafayette}{Department of Physics, Lafayette College, 730 High St, Easton, PA 18042, USA}

\newcommand{\wwu}{Department of Physics \& Astronomy, Western Washington University, Bellingham, WA 98225, USA}

\newcommand{\cfa}{Center for Astrophysics $\vert$ Harvard $\&$ Smithsonian, 60 Garden St, Cambridge, MA 02138, USA}

\newcommand{\unc}{Department of Physics and Astronomy, University of North Carolina, Chapel Hill, NC 27599, USA}

\newcommand{\ut}{Department of Astronomy, The University of Texas at Austin, Austin, TX 78712, USA}

\newcommand{\exeter}{Department of Physics and Astronomy, Stocker Road, University of Exeter, EX4 4QL, UK}

\author[0000-0002-8047-1982]{Alejandro N\'u{\~n}ez}
\altaffiliation{NSF MPS-Ascend Postdoctoral Research Fellow}
\affil{\columbia}

\author[0000-0001-7077-3664]{Marcel A.~Ag\"{u}eros}
\affil{\columbia},\affil{\LAB}

\author[0000-0002-2792-134X]{Jason L.~Curtis}
\affiliation{\columbia}

\author[0000-0001-6914-7797]{Kevin R.~Covey}
\affiliation{\wwu}

\author[0000-0001-7371-2832]{Stephanie T.\ Douglas}
\affiliation{\lafayette}

\author[0009-0008-7414-8039]{Sabine R.~Chu}
\affiliation{\columbia}

\author[0000-0002-8922-825X]{Stanislav DeLaurentiis}
\affiliation{\columbia}

\author{Minzhi (Luna) Wang}
\affiliation{\columbia}

\author[0000-0002-0210-2276]{Jeremy J.\ Drake}
\affiliation{\cfa}

\begin{abstract}

% max 250 words
Low-mass ($\lesssim$1.2 \Msun) main-sequence stars lose angular momentum over time, leading to a decrease in their magnetic activity. The details of this rotation--activity relation remain poorly understood, however. Using observations of members of the $\approx$700 Myr-old Praesepe and Hyades open clusters, we aim to characterize the rotation--activity relation for different tracers of activity at this age. To complement published data, we obtained new optical spectra for 250 Praesepe stars, new X-ray detections for ten, and new rotation periods for 28. These numbers for Hyads are 131, 23, and 137, respectively. The latter increases the number of Hyads with periods by 50\%. We used these data to measure the fractional \halpha\ and X-ray luminosities, \LLH\ and \LLX, and to calculate Rossby numbers \Ro. We found that at $\approx$700 Myr almost all M dwarfs exhibit \halpha\ emission, with binaries having the same overall color--\halpha\ equivalent width distribution as single stars. In the \Ro--\LLH\ plane, unsaturated single stars follow a power-law with index $\beta = -5.9$$\pm$$0.8$ for \Ro~$>0.3$. In the \Ro--\LLX\ plane, we see evidence for supersaturation for single stars with \Ro\ $\lesssim$ 0.01, following a power-law with index $\beta_\mathrm{sup} = 0.5^{+0.2}_{-0.1}$, supporting the hypothesis that the coronae of these stars are being centrifugally stripped. We found that the critical \Ro\ value at which activity saturates is smaller for \LLX\ than for \LLH. Finally, we observed an almost 1:1 relation between \LLH\ and \LLX, suggesting that both the corona and the chromosphere experience similar magnetic heating.

\end{abstract}


%%%%%%%%%%%%%%%%%%%%%%%%%%%%%%%%%%%%%%%%%%%%%%%%%%%%%%%%%%%%%%
\section{Introduction}

The magnetic field of a low-mass, main-sequence star ($\lesssim$1.2 \Msun) is generated by a complex dynamo, which arises from differential rotation and radial convective motions in the outer convective envelope \citep[cf.~review in][]{Fan2021}. The magnetic field injects energy into the stellar atmosphere \citep[e.g.,][]{Vernazza1981, Nelson2013}, and produces magnetized winds. Through these winds, the star loses angular momentum, and this weakens the magnetic dynamo that generates the magnetic field \citep{Parker1993}. 

While this picture is widely accepted, many unknowns remain about the nature of stellar magnetic activity and its connection to rotation. For instance, the intensity of different activity indicators %(e.g., X-ray emission and optical and ultraviolet spectral lines) 
is commensurate to the amount of heat generated by the magnetic activity \citep{Schrijver1989, Pevtsov2003, Guedel2004, Reiners2007, Reiners2012a}, yet how much energy is injected at different atmospheric heights is not fully understood \citep[e.g.,][]{Stelzer2013, Stelzer2016, Richey-Yowell2019}. In addition, the mechanism generating magnetic fields in solar-type stars has long been thought to rely on the existence of the tachocline, the shear layer between the radiative interior and the outer convective region \citep{Ossendrijver2003, Miesch2005}, but fully convective stars, which lack a tachocline, nonetheless show strong magnetic activity  \citep[e.g.,][]{Reiners2007, Wright2018}. 

The typical approach to quantifying the relationship between activity and rotation is to examine the behavior of the fractional luminosity of an activity indicator, i.e., the luminosity of the activity indicator divided by the bolometric luminosity \Lbol\ of the star, as a function of the Rossby number \Ro, defined as the rotation period \prot\ divided by the convective turnover time $\tau$ \citep[e.g.,][]{Noyes1984, Randich1998, Cook2014}.

X-rays, which originate in the coronae of low-mass stars \citep{Vaiana1981}, and \halpha\ emission, which originates in their chromospheres \citep{Campbell1983}, are well-known activity indicators. In the \Ro--\LLX\ and \Ro--\LLH\ planes, low-mass stars are generally in one of two regimes. For large \Ro\ ($\gtrsim$0.1), activity decreases as \Ro\ increases (i.e., as \prot\ increases). By contrast, for small \Ro, activity is independent of \Ro\ and appears to saturate at a given level, which differs for each activity indicator \citep[e.g.,][]{Stauffer1994a, Randich2000b, Pizzolato2003, Wright2011, Nunez2015}. Still, many details remain unexplained. For example, it is unclear which magnetic properties define the power-law relation in the unsaturated regime, or what sets the \Ro\ value separating unsaturated from saturated stars.

Some studies have also found that at very small \Ro\ ($\lesssim$0.01), X-ray activity levels decrease once again. This is the so-called supersaturated regime \citep{Randich1996, Stauffer1997, Jeffries2011, Argiroffi2016, Thiemann2020, Alexander2012, Cook2014}. Several hypotheses exist to explain the transition between the saturated and supersaturated regimes \citep[cf.~discussion in][]{Wright2011}. For example, \citet{Vilhu1984} suggested that the fraction of the stellar surface covered by star spots reaches a maximum at some (high) magnetic field strength, thus effectively capping the amount of activity. \citet{Solanki1997} proposed instead that the magnetic field in ultrafast-rotating stars gets concentrated near the poles, thus decreasing the amount of activity elsewhere \citep[see also][]{Stepien2001}. In these scenarios, supersaturation would affect all of the stellar atmosphere, and therefore be observed in any tracer of magnetic activity.

Alternatively, \citet{Jardine1999} developed the idea of coronal stripping, in which the outermost layers of the corona are centrifugally lost in ultrafast rotators \citep[see also][]{Jardine2004}. In this scenario, only the corona, itself the outermost atmospheric layer, would display supersaturation. Observationally, \citet{Marsden2009} found tentative evidence of coronal supersaturation in a sample of $\approx$30-Myr-old solar-type stars that showed no evidence of chromospheric supersaturation. 

The samples used in the studies described above are usually heterogeneous, including both young stars from open clusters and field-age stars \citep[e.g.,][]{Jeffries2011, Wright2011, Stelzer2016}. They may also include binaries, which may have very different evolutionary histories from their single counterparts, thanks to possible magnetic interactions with their close stellar---or sub-stellar---companions \citep[e.g.,][]{Stelzer2001, Wright2011}. Or they are restricted to solar-type stars, leaving gaps in our understanding of the magnetic behavior of their lower mass, fully convective cousins. Indeed, persuasive evidence for coronal supersaturation has only been found in G and K dwarfs \citep[e.g.,][]{Prosser1996, Stauffer1997b, Pizzolato2003, Jackson2010}, and only tentative evidence for M dwarfs \citep[e.g.][]{James2000, Reiners2010, Nunez2022}.

Open clusters are ideal laboratories for placing observational constraints on the rotation--activity relation. Stars from the same open cluster have both the same age and metallicity, allowing for a more robust characterization of the \Ro--\LLX\ or \Ro--\LLH\ planes. This paper is the fifth in our study of rotation and activity in the $\approx$700-Myr-old Praesepe and Hyades open clusters, which form a crucial bridge between the studies of very young groups of stars and those with field ages.

Two of our previous papers are especially relevant to this one. In \citet[][hereafter Paper~II]{Douglas2014}, we combined new and archival optical spectra with literature \prot\ and X-ray data to show that chromospheric and coronal activity depend differently on \prot. In \citet[][Paper~IV]{Nunez2022}, we presented an in-depth analysis of X-ray activity and rotation in both clusters, benefiting from the large number of \prot\ measurements published by \citet{Douglas2016, Douglas2017, Douglas2019} and \citet{Rampalli2022}.

In this paper, we present our analysis of hundreds of low-mass members of the two clusters with \LLH, \LLX, and \Ro\ measurements. We begin by describing updates to our membership catalogs in \autoref{sec_mem}, our optical spectroscopic data in \autoref{sec_spec}, our X-ray data in \autoref{sec_xrays}, and our photometric light-curves and \prot\ measurements in \autoref{sec_prots}. We derive several parameters for the cluster stars in \autoref{sec_derived}. We present our results in \autoref{sec_results} and conclude in \autoref{sec_conclusions}.

%%%%%%%%%%%%%%%%%%%%%%%%%%%%%%%%%%%%%%%%%%%%%%%%%%%%%%%%%%%%%%
\section{Membership Updates}
\label{sec_mem}

We adopted the original cluster membership catalog presented in Table 2 of \citetalias{Nunez2022} for Praesepe and Hyades stars, updating the Gaia data to the values  published in the Data Release 3 \citep[DR3;][]{GaiaDR3}. For Praesepe, the catalog has 1739 members, 539 of which are candidate or confirmed binaries. For Hyades, the numbers are 1315 and 298, respectively.

We updated the catalog entry for the Hyad \object{2MASS J05301288+2038486} to reflect the fact that there are two Gaia DR3 sources associated with it, one of which was not included in the Gaia Data Release 2 \citep[DR2;][]{GaiaDR2}.\footnote{The two objects are \object[Gaia DR3 3402090466142958464]{3402090466142958464} ($G$ = 11.38 mag, $\varpi$ = 13.30$\pm$0.02 mas, $\mu_\alpha \cos \delta$ = 26.00$\pm$0.03 mas yr$^{-1}$, $\mu_\delta$ = $-$21.02$\pm$0.02 mas yr$^{-1}$, RV = 0.88$\pm$0.28 km s$^{-1}$) and \object[Gaia DR3 3402090466140560128]{3402090466140560128} ($G$ = 14.84 mag, $\varpi$ = 13.28$\pm$0.21 mas, $\mu_\alpha \cos \delta$ = 27.72$\pm$0.27 mas yr$^{-1}$, $\mu_\delta$ = $-$19.31$\pm$0.17 mas yr$^{-1}$, RV = $-$0.70$\pm$6.32 km s$^{-1}$); the latter one is not in DR2.} Whether these two DR3 sources are gravitationally bound or unassociated remains to be confirmed, but for the purpose of this study, we categorize the star as a binary and change its binary flag from 0 (not binary) to 1 (candidate binary). 
\input{tbl_measurements}

We also updated the binary flag for Hyads \object{2MASS J02594633+3855363} and \object[2MASS J04461522+1846294]{J04461522+1846294} from 0 to 1. Our TESS light curve analysis (see \autoref{sec_prots}) revealed multiple periodicity in these two stars. As such, we considered them candidate binaries for this study (see Section~\ref{sec_prots} for more details). The updated catalog for Hyades therefore now has 1312 single members and 301 candidate or confirmed binaries.

We present our updated catalog in \autoref{tbl_measurements}, with our adopted name for each star in Column 1. Columns 2 and 3 include 2MASS \citep{2mass} and Gaia DR3 designations. Column 4 identifies the cluster to which the stars belongs. Our updated binary flags are given in Column 5.  In the following Sections we describe the rest of the columns in the table.
\input{tbl_MDMlog}

%%%%%%%%%%%%%%%%%%%%%%%%%%%%%%%%%%%%%%%%%%%%%%%%%%%%%%%%%%%%%%
\section{Optical Spectroscopy}\label{sec_spec}
\begin{figure*}
\centerline{\includegraphics[width=.99\linewidth]{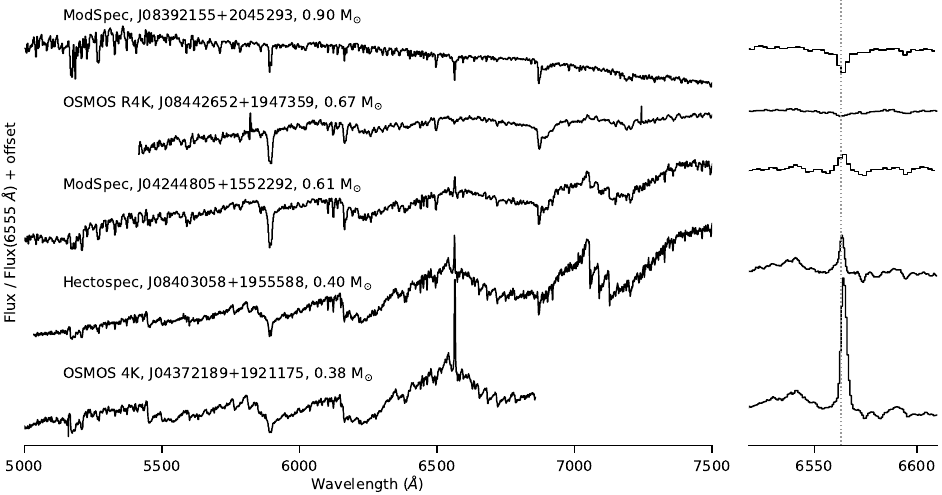}}
\vspace{-.2cm}
\caption{Five representative Praesepe and Hyades spectra obtained with the 2.4-m Hiltner telescope at MDM (Modspec and OSMOS spectrographs) and the MMT Observatory (Hectospec). Each spectrum is labeled with the instrument used, 2MASS designation of the target, and stellar mass $m$, and is normalized to the flux at 6555~\AA. The right panel shows a close-up of the \halpha\ line. The vertical dotted line indicates the center of the \halpha\ line.}
\label{fig_spectra}
\end{figure*}

In \citetalias{Douglas2014}, we presented new spectra for 130 Hyads and 390 Praesepe members obtained with the MDM Observatory 2.4-m Hiltner telescope and the WIYN 3.5-m telescope, both on Kitt Peak, AZ, and with the Magellan 6.5-m Clay telescope, Las Campanas Observatory, Chile. To these, we added archival spectra from \citet{Allen1995}, \citet{Stauffer1997}, and \citet{Kafka2004, Kafka2006}, and from the Sloan Digital Sky Survey \citep[SDSS;][]{york00} archive. The resulting spectroscopic sample  contained 720 spectra for 516 Praesepe members and 139 spectra for 130 Hyads. 

\subsection{New MDM Observations}\label{MDM}
We obtained additional spectra of Praesepe and Hyades stars over the course of 19 observing runs between 2014 Nov and 2023 Mar with the Modular Spectrograph (ModSpec) and the Ohio State Multi-Object Spectrograph (OSMOS) on the MDM Observatory 2.4-m Hiltner telescope (see \autoref{mdmstats}). We configured ModSpec to have a wavelength coverage of 4500--7500 \AA\ with $\approx$1.8 \AA\ sampling and R $\approx$ 3600, which is the same configuration we used in \citetalias{Douglas2014}. 

With OSMOS, we used the blue 4K detector (OSMOS 4K) with a 1.2\asec\ inner slit, for an approximate wavelength coverage of 4000--6800 \AA, $\approx$0.7 \AA\ sampling, R~$\approx$~9300, and peak efficiency at 6400~\AA. We also used the red 4K detector (OSMOS R4K) with an OG-530 longpass filter and 1.2\asec\ center slit, for an approximate wavelength coverage of 5500--10000 \AA, $\approx$1.3 \AA\ sampling, R~$\approx$~5000, and peak efficiency near 9000 \AA.

MDM spectra obtained before 2021 were reduced with a script written in PyRAF,\footnote{\url{https://pypi.org/project/pyraf/}} the Python-based command language for the Image Reduction and Analysis Facility \citep[IRAF;][]{IRAF}. Spectra obtained in 2021 and later were processed with the Python package {\tt PypeIt} \citep[Version 1.10.1.dev3+g52d10edd;][]{Prochaska2020, Prochaska2020zndo}. We tested the agreement between the PyRAF and {\tt PypeIt} pipelines by reducing a small sample of raw OSMOS images with both pipelines; most of these data were for stars with \halpha\ absorption. We then measured the \halpha\ equivalent width (see \autoref{halpha}) in all spectra. The difference in measurements between the pipelines were $<$10\%.

All the spectra were trimmed, overscan- and bias-corrected, cleaned of cosmic rays, flat-fielded, extracted, dispersion-corrected, and flux-calibrated. Excluding poor quality spectra (e.g., $S$/$N$ $\lesssim$~5 or acquisition imperfections; 12\% of Praesepe spectra and 8\% of Hyades spectra), we collected 454 new spectra for 153 Praesepe members and 231 for 209 Hyads. The median $S$/$N$ for these spectra is 76 at \halpha.

Spectra for four stars observed at MDM are shown in \autoref{fig_spectra} for illustrative purposes. The MDM (and WIYN) spectra from \citetalias{Douglas2014} and this work are available online.\footnote{Available at \url{https://doi.org/10.7916/8ag4-4c53}.\label{fn:commons}} 

\subsection{New MMT Observations}\label{MMT}
We obtained additional spectra of Praesepe stars with the multiobject spectrograph Hectospec \citep{Fabricant2005} on the MMT 6.5-m telescope, Mt.~Hopkins, AZ. We used two fiber configurations over the course of two consecutive nights (2015 Nov 21-22). The first configuration was centered near $\alpha=08^{\mathrm{h}}41^{\mathrm{m}}10^{\mathrm{s}}$, $\delta=+20^{\circ}09'59\farcs1$ (J2000) and targeted 57 Praesepe stars. The second configuration was centered near $\alpha=08^{\mathrm{h}}41^{\mathrm{m}}36^{\mathrm{s}}$, $\delta=+19^{\circ}03'50\farcs5$ (J2000) and targeted 50 additional Praesepe stars.

We used the 600 line grating centered at 6300~\AA, which results in an approximate wavelength coverage of 5030-7540 \AA, and give $R \approx 11,000$ at \halpha. Our targets had 14.9 $ < G < $ 19.7 mag, and our integration times were 3600 s (first night) and 5400 s (second night) with the first configuration, and 4500 s (second night) with the second configuration. After excluding spectra with $S$/$N$~$\lesssim$~5, we have 126 MMT spectra for 47 Praesepe stars. The median $S$/$N$ at \halpha\ is 48. All our MMT spectra are also available online.

The data were reduced automatically by the Smithsonian Astrophysical Observatory Telescope Data Center using the HSRED v2.0 pipeline. HSRED performs the basic reduction tasks: bias subtraction, flat-fielding, arc calibration, and sky subtraction.\footnote{See \url{http://mmto.org/~rcool/hsred/index.html} for a  description of HSRED.} A Hectospec spectrum is included in \autoref{fig_spectra} for illustrative purposes. The MMT spectra are available online.\footref{fn:commons}

\input{tbl_Xobs}

\subsection{New Archival Spectroscopy}\label{arch}

In \citetalias{Douglas2014}, we found SDSS spectra for 66 Praesepe stars (as of 2013 Feb 14). We repeated this search using the SDSS Science Archive Server.\footnote{\url{https://dr16.sdss.org/home}} SDSS spectra are sky-subtracted, corrected for telluric absorption, spectrophotometrically calibrated, and calibrated to heliocentric vacuum wavelengths. The wavelength coverage is 3800-9200 \AA\ with R $\approx$ 4300 at \halpha. After excluding those with $S$/$N$ $\lesssim$~5, we found SDSS spectra for 102 Praesepe stars and 16 Hyads (as of 2023 May 21).

We also searched for spectra in the the Large Sky Area Multi-Object Fiber Spectroscopic Telescope (LAMOST) Data Release 8 catalog.\footnote{\url{http://dr8.lamost.org/v2/}} These spectra are flux- and wavelength-calibrated and sky-subtracted, and cover %the wavelength range of 
3690-9100~\AA\ with R~$= 1800$ at 5500~\AA. After excluding those with $S$/$N$ $\lesssim$~5, we found 873 LAMOST spectra for 324 Praesepe members and 535 for 252 Hyads. This includes 108 stars in Praesepe and 146 in the Hyades that did not have any previously available spectra. 

Finally, J.~Stauffer (2014, priv.~comm.) shared 12 spectra of Hyads obtained as part of the \citet{Stauffer1997} survey of the cluster. In \citetalias{Douglas2014} we used 10 of these spectra to compare our equivalent width measurements to those of \citet{Stauffer1997};  here we include all 12 in our spectroscopic sample.

With our newly obtained spectra and newly found archival spectra, we now have a total of 2216 good quality (signal-to-noise ratio $S$/$N$ $\gtrsim$~5) spectra for 879 Praesepe members; for the Hyades, the numbers are 943 and 565, respectively. Ninety-four of the Praesepe stars and 12 Hyads have five or more spectra.

%%%%%%%%%%%%%%%%%%%%%%%%%%%%%%%%%%%%%%%%%%%%%%%%%%%%%%%%%%%%%%
\section{X-ray data}\label{sec_xrays}

\citetalias{Nunez2022} explains in detail the origin of the X-ray data for our Praesepe and Hyades stars. Briefly: we consolidated X-ray detections from the R\"ontgen Satellite (ROSAT), the Chandra X-ray Observatory (Chandra), the Neil Gehrels Swift Observatory (Swift), and the X-ray Multi-Mirror Mission Newton (XMM). For faint X-ray sources, we converted instrumental counts to unabsorbed energy fluxes \fx\ using WebPIMMS\footnote{\url{https://heasarc.gsfc.nasa.gov/cgi-bin/Tools/w3pimms/w3pimms.pl}}, and for bright X-ray sources, we performed spectral analyses to extract unabsorbed \fx. For sources with flares in their X-ray light curves, we removed counts from the flare events before calculating \fx\ to obtain a more representative measurement of the quiescent X-ray activity level. We also homogenized all the fluxes to the 0.1--2.4 keV energy band. 

Finally, for stars with more than one X-ray detection, we calculated the error-weighted average of the \fx\ values and adopted it as the \textit{bona fide} unabsorbed \fx\ for that star. We include unabsorbed \fx\ values and their standard deviations (1$\sigma$ uncertainties) for Praesepe and Hyades stars in Columns 6 and 7 of \autoref{tbl_measurements}.\\
\input{tbl_newX.tex}
For this work, we updated our X-ray data to reflect developments since the publication of \citetalias{Nunez2022}, namely new Chandra observations, additions to the Chandra Source Catalog (CSC), and the thirteenth data release of the XMM EPIC Serendipitous Source Catalogue \citep[4XMM-DR13;][]{4XMM}.

As part of the Chandra Cool Targets program\footnote{\url{https://cxc.harvard.edu/proposer/CCTs.html}} (Proposal 20201075, PI: Ag\"ueros), we observed 17 Hyads with 14 pointings. The details of these observations are given in \autoref{chandra}. For each observation, we used the ACIS-S3 chip in Very Faint telemetry mode. We processed the raw observations with the Chandra Interactive Analysis of Observations \citep[CIAO;][]{Fruscione2006} tools.\footnote{We used CIAO v.4.14 and CALDB v.4.10.2; see section 3.2.2 in \citetalias{Nunez2022} for a full description of the data reduction.} We include in \autoref{tbl_newX} the X-ray data for 13 of the targeted stars; four were undetected. We note that \autoref{tbl_newX} in this work is an addendum to Table 3 in \citetalias{Nunez2022}.

In addition, two Hyads were added to the CSC following the release of version 2.1 starting in late 2022. Data for these two new X-ray detections are also included in \autoref{tbl_newX}. 

Lastly, we found nine Praesepe low-mass members in 4XMM-DR13 that had no previous X-ray detections and one more that had ROSAT and Swift X-ray detections. We also found four Hyads with no previous X-ray detections, and four Hyads that only had a ROSAT X-ray detection. Data for these 18 4XMM-DR13 detections are included in \autoref{tbl_newX}.

%%%%%%%%%%%%%%%%%%%%%%%%%%%%%%%%%%%%%%%%%%%%%%%%%%%%%%%%%%%%%%
\section{Rotation Period Measurements}\label{sec_prots}

\begin{figure*}[t]
\centerline{\includegraphics{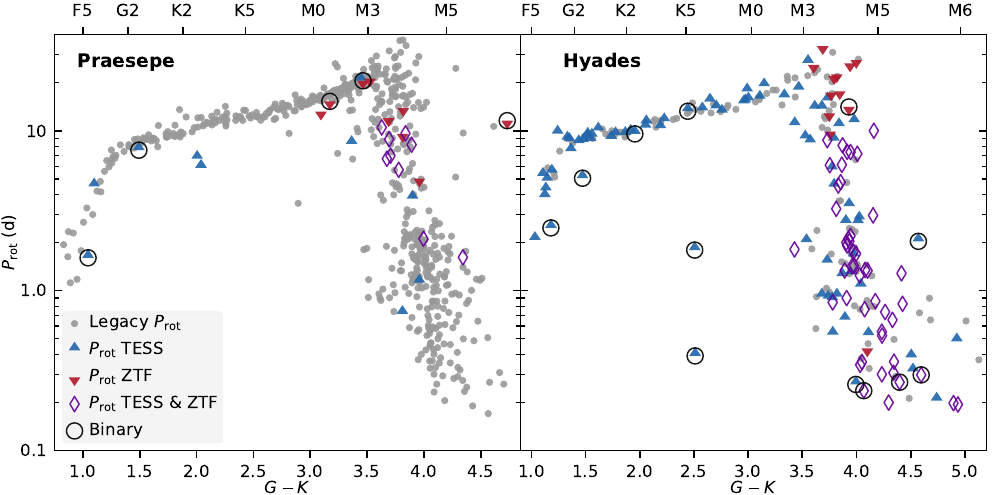}}
\vspace{-.2cm}
\caption{\prot\ vs.~\gminusk\  for Praesepe (left panel) and Hyades (right panel) stars. Gray circles indicate single stars with existing K2 \prot\ and other archival data collected in \citet{Rampalli2022} and \citet{Douglas2019} for Praesepe and the Hyades, respectively. Blue up triangles, red down triangles, and purple diamonds indicate stars with new \prot\ values from TESS, ZTF, or both, respectively. Black circles highlight new \prot\ values for known and candidate binaries.}
\label{fig_newprots}
\end{figure*}

The bulk of our rotational data for Praesepe and the Hyades came from the catalogs published in \citet{Rampalli2022} and \citet{Douglas2019}, respectively. These catalogs consolidated \prot\ measurements made from light curves obtained by ground-based photometric surveys and by K2 \citep{howell2014}. For Praesepe, we have 1052 members with these legacy \prot\ measurements; for the Hyades, the number is 233.

More recent observations of the two clusters with the Transiting Exoplanet Survey Satellite \citep[TESS;][]{Ricker2015}, and continuing observations of the clusters with the Zwicky Transient Facility \citep[ZTF;][]{masci2018}, provided an opportunity to add to these totals. We showcase in \appref{app:lcexample} the differing qualities of TESS and ZTF data and the benefit of using them together to extract more reliable rotation periods. Accordingly, we searched for light curves for stars in both clusters for which we have an optical spectrum and/or an X-ray detection.\footnote{Completing the \prot\ census for all of the stars in either cluster, i.e., to obtain new \prot\ values for stars without magnetic activity measurements, is beyond the scope of this paper.} 

Our procedure for TESS followed the strategy employed in a number of recent studies \citep[e.g.,][]{mcdivitt2022}. We downloaded 40$\times$40 pixel cutouts from the available full frame images using \texttt{TESScut} \citep{tesscut}, hosted by MAST.\footnote{All the TESS data used in this paper can be found in MAST: \dataset[10.17909/0cp4-2j79]{http://dx.doi.org/10.17909/0cp4-2j79}.} We extracted light curves for the pixel closest to each target using Casual Pixel Modeling \citep{wang2016} as implemented in the package \texttt{unpopular}  \citep{Hattori2022}. 

Using the interactive program \texttt{tesscheck},\footnote{\url{https://github.com/SPOT-FFI/tess_check}} we then inspected the TESS light curves for each star, selected the optimal subset of sectors to search for a rotational signature, and measured the period using Lomb--Scargle periodograms \citep{press1989}. The visual inspection allowed us to catch uncorrected systematics that can introduce spurious signals into the periodogram and to flag and correct cases where the periodogram favors the half-period harmonic. We measured periods for 19 Praesepe stars and 125 Hyads using TESS data.

Our ZTF procedure used the approach developed by \citet{Curtis2020} to analyze Palomar Transient Factory \citep{Law2010} data for the 2.7-Gyr-old cluster Ruprecht~147. We downloaded 8$\amin$$\times$8$\amin$ cutout images from IPAC\footnote{All the ZTF data used in this paper can be found in IPAC: \dataset[10.26131/IRSA539]{http://dx.doi.org/10.26131/IRSA539}.} \citep{ZTFdata} and used simple aperture photometry to extract light curves for our targets and neighboring reference stars identified with Gaia. We corrected the light curves for systematics using the median-combined normalized light curves for reference stars. We inspected the resulting light curves, isolated the segment showing the cleanest periodic variability, and measured the period using Lomb--Scargle periodograms. We measured periods for 18 Praesepe stars and 59 Hyads using ZTF data.

In total, we have new \prot\ measurements for 28 Praesepe stars and 137 Hyads; 56 of these 165 stars have periods determined from both TESS and ZTF. For the Hyades, we have increased the sample of cluster members with known \prot\ by $\approx$50\%, bringing the total to 370 stars. \autoref{fig_newprots} highlights our new \prot\ measurements against the background of legacy \prot\ for both clusters. In \autoref{tbl_measurements}, we include the \prot\ data for Praesepe and Hyades members in Column 8, and we identify the source of the \prot\ measurement in Column 9.

We measured \prot~=~0.39~d from the TESS data for \object{2MASS J05301288+2038486}. This is an unusually short period for a star of its color; the other single stars with \gminusk\ $\approx$ 2.5 mag in \autoref{fig_newprots} have \prot\ between 10 and 20 d. As mentioned in \autoref{sec_mem}, we found two Gaia DR3 sources within $<$2$\asec$ of each other and associated with this 2MASS source. Given this, we consider \object{2MASS J05301288+2038486} a candidate binary and flag it as such in \autoref{fig_newprots}.

\begin{figure}[t]
\centerline{\includegraphics{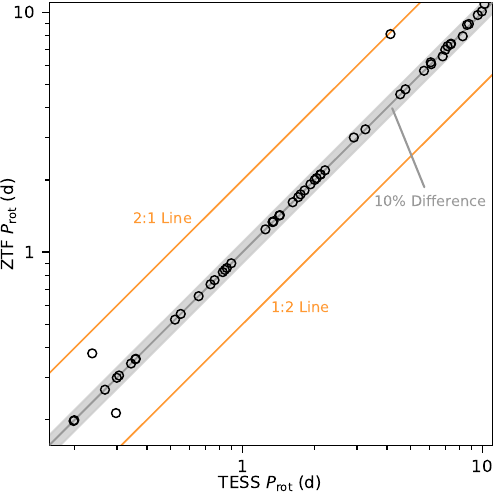}}
\vspace{-.2cm}
\caption{\prot\ from ZTF vs.~\prot\ from TESS for the 56 Praesepe and the Hyades stars for which both surveys yielded periods. The gray line is the 1:1 relation, and the gray area the $\pm$10\% difference range. The orange lines indicate the 1:2 and 2:1 harmonic lines. The two stars outside the 10\% difference range have indications of multiple periodicity (see \autoref{app:lcs}); we adopted the TESS \prot\ and flag them as candidate binaries. For the star falling on the 2:1 harmonic line, we adopted the ZTF \prot.}
\label{fig_protcomp}
\end{figure}

\autoref{fig_protcomp} compares our TESS- and ZTF-derived \prot\ for the 56 cluster stars for which we have both. For 53 of the 56, the two \prot\ disagree by $<$4\%. Of the three stars for which the disagreement is larger, two have TESS light curves that suggest multiple periods (see \appref{app:lcdisagreement}). Indeed, for these two stars, \object{2MASS J02594633+3855363} and \object[2MASS J04461522+1846294]{J04461522+1846294}, the Gaia re-normalized unit weight error (RUWE) values are 5.0 and 3.4, respectively. As discussed in \citetalias{Nunez2022}, stars with RUWE~$>$~1.4 have a high probability of being unresolved binaries \citep[e.g.,][]{Deacon2020, Ziegler2020, Kervella2022}. For these two stars, we adopted the TESS \prot, assigned a binary flag = 1 (indicating they are candidate binaries), and flagged them as such in \autoref{fig_newprots}.

For the third star, \object{2MASS J08412772+2103409}, the TESS \prot\ (4.1 d) is the half harmonic of the ZTF \prot\ (8.2~d). We adopted the ZTF \prot\ for this star.

%%%%%%%%%%%%%%%%%%%%%%%%%%%%%%%%%%%%%%%%%%%%%%%%%%%%%%%%%%%%%%
\section{Other measurements and Derived Quantities}\label{sec_derived}

\subsection{H$\alpha$ Equivalent Width Measurements}\label{halpha}

We measured the equivalent width (EW) of the \halpha\ Balmer line in all our optical spectra, both newly acquired and archival (see \autoref{sec_spec}). For this purpose, we used the tool \texttt{PHEW} \citep{PHEW}, which automates the EW measurement using \texttt{PySpecKit} \citep{Ginsburg2011} to fit a Voigt profile to the \halpha\ line. We interactively defined continuum regions to either side of the \halpha\ line in each spectrum, each between 5 and 35 \AA\ in length. 

\texttt{PHEW} performs 1000 Monte Carlo iterations by re-sampling the flux measurements within the flux uncertainties, or, if flux uncertainties are unavailable, by adding Gaussian noise to the flux spectrum. It then calculates the standard deviation of the 1000 EWs, which we adopted as the 1$\sigma$ EW uncertainty. We extracted the noise for each point from a Gaussian with width equal to the associated uncertainty at that point. If a flux spectrum lacked an associated uncertainty spectrum, we instead extracted the noise from a Gaussian with width equal to the standard deviation of the flux in the continuum regions defined for that spectrum.

If a star had multiple spectra available, we adopted the error-weighted mean EW (and the weighted mean standard error) as the representative EW value (and 1$\sigma$ uncertainty) for that star. Columns 10 and 11 in \autoref{tbl_measurements} include our measured EW and its 1$\sigma$ uncertainty, respectively. Negative EWs indicate emission, and an EW value of zero indicates that the spectrum for the star does not display a measurable \halpha\ feature at that spectral resolution. Column 12 indicates the number of spectra we used to calculate each star's EW value. The output figures from \texttt{PHEW} showing our EW measurements are available online.\footref{fn:commons}

\begin{figure*}[ht]
\centerline{\includegraphics{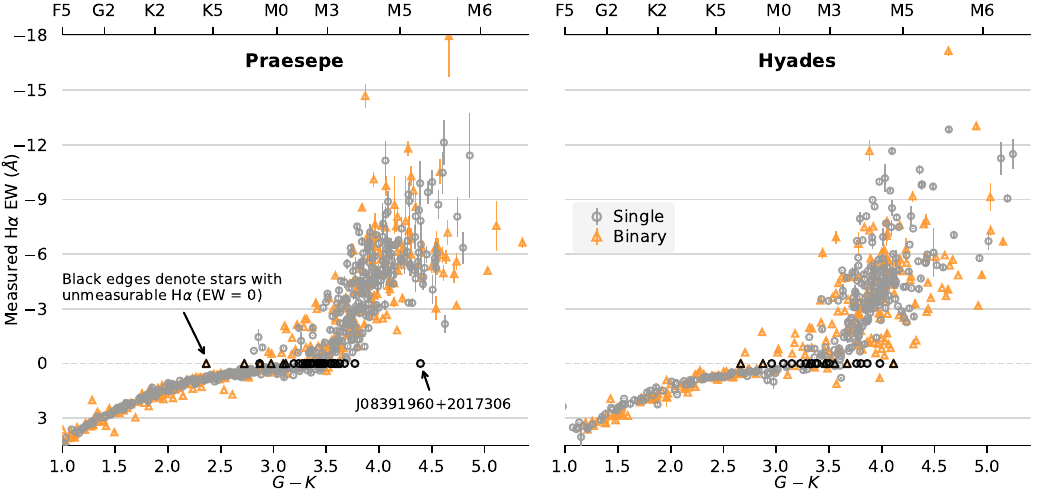}}
\vspace{-0.2cm}
\caption{Measured \halpha\ EW vs.~($G - K$) for Praesepe (left panel) and Hyades (right panel) members. Single stars are indicated with gray circles and binaries with orange triangles. Most of the EW error bars are smaller than the symbols. For clarity, we excluded from the right panel three outlier stars, one with \gminusk\ $>$ 5.5 mag and two with \halpha\ EW $< -18$ \AA. We also excluded stars with \gminusk\ $<$ 1.0 (spectral types earlier than $\approx$F5), as they are not relevant to our analysis. Black symbols indicate stars for which \halpha\ is immeasurable in our spectra, and for which we set EW = 0. We consider one of these stars, annotated with its 2MASS designation, to be a potential non-member based on its unusual inactivity (see \autoref{halpha}). The EWs shown here were not corrected for the quiescent \halpha\ absorption present in these stars.}
\label{fig_ew_bins}
\end{figure*}

\autoref{fig_ew_bins} shows our EW measurements as a function of \gminusk\ for single and binary members (gray circles and orange triangles, respectively) of Praesepe (left panel) and of the Hyades (right panel). To better visualize the overall pattern, we omitted from the figure three Hyades outliers, one with \gminusk\ $>$ 5.5 mag and two with EW $< -18$ \AA. We also excluded stars with \gminusk\ $<$ 1 (spectral type earlier than $\approx$F5) from the figure, as their lack of significant convective envelopes implies a different rotational evolution than the solar-like stars we focus on. However, we include in \autoref{tbl_measurements} values for all stars with at least one spectrum, regardless of spectral type.

In \autoref{fig_ew_bins}, we also highlight with black symbols stars for which we find no measurable \halpha\ (EW = 0~\AA). Among these stars is \object{2MASS J08391960+2017306}, a Praesepe star with \gminusk\ = 4.4 mag. All of its M5-M6 cousins exhibit some level of \halpha\ emission, which makes its \halpha\ inactivity unusual. In \citetalias{Nunez2022}, we considered this star a plausible Praesepe member based on its \citet{Kraus2007} membership probability of $\approx$70\%. However, none of the Gaia-based membership studies included in \citetalias{Nunez2022} considered it a cluster member, as its Gaia data do not include parallax and proper motion information (as of DR3). We therefore believe this star to be a likely contaminant in our membership catalog for Praesepe.\footnote{We have no X-ray detection or period for this star, so it does not appear elsewhere in our analysis.}

\subsection{H$\alpha$ Relative to Quiescence}\label{sec_relativeEW}

We corrected our measured EW values for each star to account for the quiescent photospheric \halpha\ absorption naturally present in low-mass stars \citep[cf.~discussion for M dwarfs in][]{Stauffer1986}. As stars become more magnetically active, the line fills in and eventually transitions to emission. To report more accurately the level of chromospheric activity, we therefore need to consider this quiescent absorption level, which is a function of stellar mass $m$.

We used the empirical model of \citet{Newton2017}, valid for stars with $m$ < 0.8 \Msun, to calculate the quiescent photospheric absorption EW for cluster stars in that $m$ range.\footnote{In principle, stars with $m$ > 0.8 \Msun\ also exhibit quiescent photospheric \halpha\ absorption. However, the main focus of our study is on stars with \halpha\ in emission, and none of our stars with spectra and $m$ > 0.8 \Msun\ fall in that category.} We then determined the \emph{relative} EW by subtracting the quiescent EW from our measured EW. Column 13 in \autoref{tbl_measurements} indicates the relative EW value for each star with a measured EW and within the $m$ range of the empirical model.

\subsection{The $\chi$ Factor and \LLH}\label{llh}
To obtain \LLH\ for stars with \halpha\ in emission, we used the relation
\begin{equation}\label{eq:llh}
  \frac{L_{\mathrm{H}\alpha}}{L_{\mathrm{bol}}} = -\mathrm{EW}_{\mathrm{H}\alpha}\ \chi,
\end{equation}
where EW$_{\mathrm{H}_{\alpha}}$ is the relative \halpha\ EW calculated in \autoref{sec_relativeEW}, and $\chi$ is the ratio of the continuum flux near the \halpha\ line and of the apparent bolometric flux.

In \citetalias{Douglas2014}, we presented several empirical $\chi$--photometric color relations based on PHOENIX ACES model spectra \citep{Husser2013}. We measured $\chi$ in the model spectra with surface gravity log($g$) = 5.0, solar metallicity, and in the effective temperature (\teff) range 2500--5200 K. For this work, we extended this calculation of $\chi$ to include the \teff\ range 2300--6500 K by following the methodology described in \citetalias{Douglas2014} (see \autoref{tbl_chi}). 

\begin{figure}
\centerline{\includegraphics{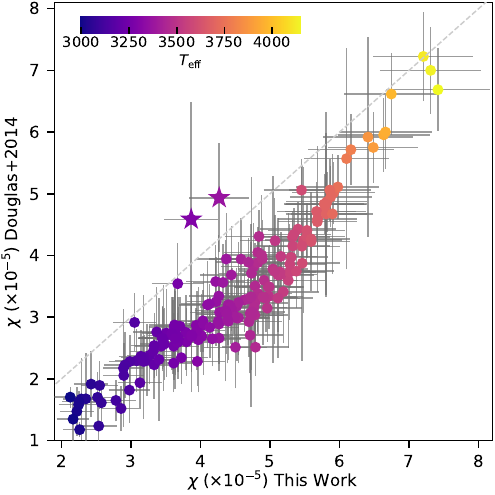}}
\vspace{-0.2cm}
\caption{$\chi$ values for single Praesepe and Hyades stars calculated in \citetalias{Douglas2014} vs.~our calculations in this work, with their respective 1$\sigma$ errors. The gray dashed line is the 1:1 relation. Stars are color-coded according to their \teff\ (as derived in this work; see Section~\ref{llh}), following the colorbar at the top left. The two stars represented with star symbols are objects for which we identified erroneous or unreliable $r^\prime$ photometry, which was used in \citetalias{Douglas2014} to estimate their $\chi$.}
\label{fig_chiscomp}
\end{figure}

To calculate $\chi$ for our cluster stars, we first derived their \teff\ using the empirical \teff--$M_G$ relation of E.~Mamajek.\footnote{Version 2022.04.16. Available at \url{http://www.pas.rochester.edu/~emamajek/EEM_dwarf_UBVIJHK_colors_Teff.txt}. Much of this table comes from \citet{Pecaut2013}.} We linearly interpolated between the $M_G$ values in the empirical relation to obtain \teff. Columns 14 and 15 in \autoref{tbl_measurements} include our derived \teff\ values and 1$\sigma$ uncertainties, respectively, for each main sequence cluster star. 

Next, we calculated $\chi$ using \teff\ by linearly interpolating between the \teff\ values in \autoref{tbl_chi}. We assumed an intrinsic 10\% error in our \teff--$\chi$ relation (identified in \citetalias{Douglas2014}) and we added this error in quadrature to produce the 1$\sigma$ of the $\chi$ values we calculated for our cluster stars. Columns 16 and 17 include our $\chi$ values and 1$\sigma$ uncertainties for each cluster star. Lastly, we applied \autoref{eq:llh} for all stars with relative \halpha\ EW and $\chi$ values to obtain \LLH.

In \autoref{fig_chiscomp}, we compare our new $\chi$ values to those in \citetalias{Douglas2014} for the sample of single Praesepe and Hyades stars in both studies. The $\chi$ values from the earlier work were derived using the log($\chi$)--($r^\prime-K$) relation. We identified two stars for which an erroneous or unreliable $r^\prime$ photometry was assigned in \citetalias{Douglas2014} (highlighted in \autoref{fig_chiscomp} with stars symbols), which explains their significant deviation from the general trend.

Our new $\chi$ values are systematically larger than those in \citetalias{Douglas2014} by a factor of $\approx$1.3. This discrepancy is mostly driven by the \teff--($r^\prime-K$) relation in \citetalias{Douglas2014}, which produces cooler \teff\ values than those derived from the E.~Mamajek table.

\input{tbl_chi}

\subsection{Bolometric Luminosities and Rossby Numbers}\label{lbol}
We used the bolometric luminosities and \Ro\ derived in \citetalias{Nunez2022} for our cluster stars. Briefly: to obtain \Lbol, we used the empirical log(\Lbol)--$M_G$ relation of E.~Mamajek.

To calculate \Ro, we first calculated $m$ using the empirical $m$--$M_G$ relation of E.~Mamajek. Next, we found the convective turnover time $\tau$ using the empirical $m$--log($\tau$) relation of \citet{Wright2018}. Finally, we computed \Ro\ = \prot/$\tau$. 

In \citetalias{Douglas2014}, we used the $m$--$M_K$ relation of \citet{Kraus2007} to calculate $m$ and the $m$--log($\tau$) relation of \citet{Wright2011} to calculate $\tau$. Compared to the \Ro\ values in \citetalias{Douglas2014}, our new \Ro\ values for the same stars are between 45\% smaller and 20\% larger, the median being 14\% smaller. The largest discrepancies are mostly due to differences in the two $m$ calculation methods and to the distances used to calculate absolute magnitudes, as \citetalias{Douglas2014} relied mostly on individual Hipparcos parallaxes or Hipparcos-derived cluster distances \citep{vanLeeuwen2009}.

%%%%%%%%%%%%%%%%%%%%%%%%%%%%%%%%%%%%%%%%%%%%%%%%%%%%%%%%%%%%%%
\section{Results and Discussion}\label{sec_results}

\subsection{Chromospheric Activity}\label{sec_ewresults}
\autoref{fig_ew_bins} shows that in both clusters, all the late F, G, and K dwarfs have converged onto a tight sequence of \halpha\ absorption (EW $>$ 0~\AA), which is independent of magnetic activity. On the other hand, most M dwarfs exhibit some level of \halpha\ emission. The transition between \halpha\ absorption and emission in the two clusters occurs essentially at the same color (corresponding to a spectral type M0-M1), suggesting that both clusters are indeed of very similar ages.

\begin{figure}[t]
\centerline{\includegraphics{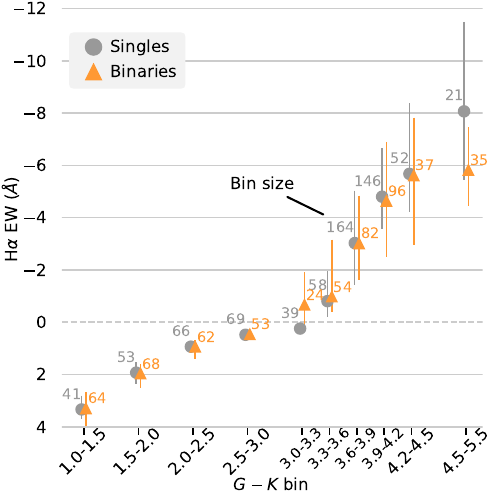}}
\vspace{-.2cm}
\caption{Color-binned median measured \halpha\ EWs for single stars (gray circles) and binaries (orange triangles) in Praesepe and Hyades, with the 16$^\mathrm{th}$ and 84$^\mathrm{th}$ percentiles represented by whiskers}. The binary sample includes all stars with RUWE > 1.4. Numbers next to symbols indicate the number of stars in each bin.
\label{fig_ewbinned}
\end{figure}

Save for a handful of late K Hyads, the binaries in \autoref{fig_ew_bins} appear to follow the same distribution as their single-star counterparts in both clusters. To compare the two distributions more carefully, we binned our EWs by \gminusk. \autoref{fig_ewbinned} shows the median EW values for single stars (gray circles) and binaries (orange triangles) in both Praesepe and the Hyades for 0.3 or 0.5 mag color bins, with their 16$^\mathrm{th}$ and 84$^\mathrm{th}$ percentiles represented by whiskers. The median EW values of single and binary stars are almost identical in most color bins, and the difference in median EW between single stars and binaries is $\lapprox$1$\sigma$ in all color bins. 

We do note a slightly higher median \halpha\ EW for binaries compared to single stars in the \gminusk\ = 3.0--3.3 mag bin (spectral types $\approx$M0--M2). However, these differences are not statistically significant, and we do not consider them to be evidence of enhanced chromospheric activity in binary systems in our sample. Lastly, the reddest color bin shows a $\approx$38\% difference in median between single and binary stars. We attribute this discrepancy to the small sample size in those bins.

\subsection{The Dependence of Chromospheric Activity on Rotation} \label{sec_rossbyresults}

To characterize the rotation--chromospheric activity relation, we followed previous authors in parametrizing the relation, in this case \Ro--\LLH, as a flat region connected to a power-law. For stars with \Ro\ $\leq \Ro_\mathrm{,sat}$, activity is saturated---i.e., constant---and equal to (\LLH)$_\mathrm{sat}$. Above $\Ro_\mathrm{,sat}$, activity declines as a power-law with index $\beta$, and is, therefore, unsaturated. Functionally, this corresponds to
\begin{equation}\label{eq:rossby}
  \frac{L_{\mathrm{H}\alpha}}{L_{\mathrm{bol}}} = \left\{
  \begin{array}{l l}
    \left(\frac{L_{\mathrm{H}\alpha}}{L_{\mathrm{bol}}}\right)_{\mathrm{sat}} & \quad \textrm{if $R_\mathrm{o}\le R_{\mathrm{o,sat}}$}\\
    C R_\mathrm{o}^{\beta} & \quad \textrm{if $R_\mathrm{o}$ > $R_{\mathrm{o,sat}}$}
  \end{array} \right.
\end{equation}
where $C$ is a constant. This model has been widely used in the literature (e.g., \citeauthor{Randich2000b} \citeyear{Randich2000b}, \citeauthor{Wright2011} \citeyear{Wright2011}, \citetalias{Douglas2014}, \citeauthor{Nunez2015} \citeyear{Nunez2015}, \citetalias{Nunez2022}).

\input{tbl_Rossbies}

\begin{figure*}[t]
\centerline{\includegraphics{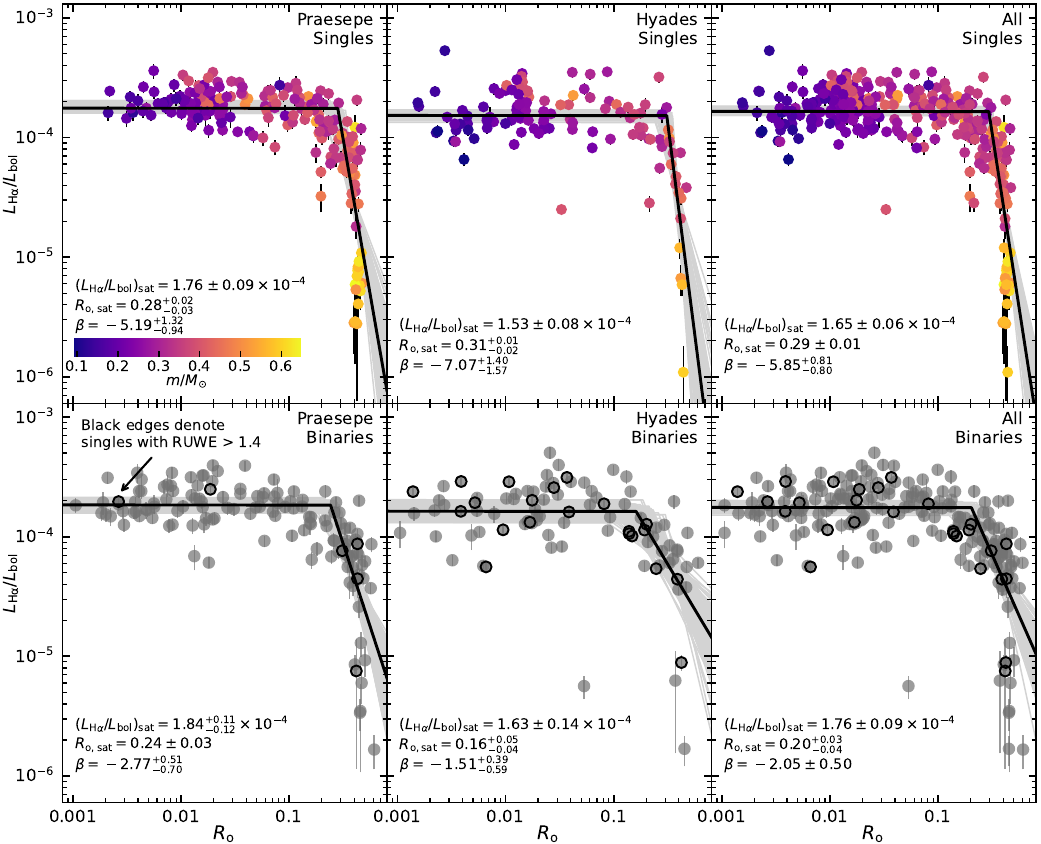}}
\caption{\LLH\ vs.~\Ro\ for Praesepe stars (left panels), for Hyads (middle), and for the two clusters combined (right). The top panels show single stars, and the bottom panels show confirmed and candidate binaries. This latter set includes nominally single stars with RUWE $>$ 1.4 (indicated with solid black circles). Single stars are color-coded by their $m$ according to the colorbar in the top left panel. The solid black line in each panel is the maximum \textit{a posteriori} fit from the MCMC algorithm, and the gray lines are 200 random samples from the posterior probability distributions. We assumed a flat saturated regime described by (\LLH)$_\mathrm{sat}$ and \Ro$_\mathrm{,sat}$, and an unsaturated regime described by a power-law of index $\beta$. The results of the fit for these three  parameters are given in each panel. We show in \autoref{app:dists} the marginalized posterior probability distributions from the MCMC analysis for each fit.}
\label{fig_rossbyHa}
\end{figure*}

We used the open-source Markov-chain Monte Carlo (MCMC) package \texttt{emcee} \citep{Foreman2013} to fit this three-parameter model to our data. Following the \texttt{emcee} implementation by \citet{Magaudda2020}, we allowed for a nuisance parameter $f$ to account for underestimated errors.\footnote{See \url{https://emcee.readthedocs.io/en/develop/user/line/}.} We assumed flat priors over each parameter and used 300 walkers, each taking 5000 steps in their MCMC chain, to infer maximum likelihood parameters. Our results are presented in \autoref{fig_rossbyHa} for several subsamples. The posterior distributions for each parameter and 2D correlations between pairs of parameters from each fit are included in a figure set in \autoref{app:dists}; 200 random samples from these distributions are shown in \autoref{fig_rossbyHa}, along with the maximum \textit{a posteriori} model. 

In Table \ref{tbl_rossbies}, we present the (\LLH)$_\mathrm{sat}$, \Ro$_\mathrm{,sat}$, and $\beta$ parameters corresponding to the maximum \textit{a posteriori} model for the six subsamples we show in \autoref{fig_rossbyHa}, and we also annotate them in each panel in the Figure. For each parameter, we assumed the 50$^\mathrm{th}$ percentile of the results to be the mean value, and the 16$^\mathrm{th}$ and 84$^\mathrm{th}$ percentiles, their approximate 1$\sigma$ uncertainties. In all cases, the nuisance parameter $f$ converged to $\approx$0.06, suggesting that our \LLH\ uncertainties are underestimated by no more than $\approx$6\%.

We applied the model in \autoref{eq:rossby} to single members separately from binary members and members with RUWE $>$ 1.4. Without a more detailed study of the characteristics of the known and candidate binaries, it is not possible to determine whether gravitational and magnetic interactions may have altered their spin-down evolution.\footnote{Gaia cannot resolve separations $\lesssim$0\farcs7 \citep{Ziegler2018}, which corresponds to a semimajor axis $a \approx 130$~au for the average Praesepe star and $\approx$35~au for the average Hyades star. Most of the candidate binaries in our sample with high RUWEs are therefore likely intermediate binaries rather than tight, tidally interacting binaries, for which $a\ \lesssim\ 0.1$ au. Still, intermediate binaries can have small enough separations ($0.1\ \lesssim\ a\ \lesssim\ 80$~au) for the binary components to have affected each other's protoplanetary disks in the first 10 Myr \citep[][]{Rebull2006, Meibom2007, Kraus2016, Messina2017}, thereby impacting their rotation--activity relation.} Furthermore, for binaries, the $\chi$ and $\tau$ parameters---ultimately derived from $M_G$ and used to calculate \LLH\ and \Ro, respectively---have dubious validity, as we expect them to be overestimated to varying degrees for binaries, the effects of which are difficult to track in our analysis. In our discussion below, we therefore distinguish between the nominally single stars and those flagged as either known or candidate binaries.

We also applied the model to members of each cluster separately and together. Combining the stars from both clusters to create a larger sample is reasonable given the very similar ages and metallicities of Praesepe and the Hyades \citep[e.g.,][]{Cummings2018, GaiaCol2018, Douglas2019}. We consider the results from the combined sample, which we nicknamed HyPra, to be most statistically meaningful. In any case, the values for the parameters obtained from applying the model to the clusters individually almost always agree to within 1$\sigma$ (and always to within 2$\sigma$).

\paragraph{The Saturated Regime}

For single HyPra stars, (\LLH)$_\mathrm{sat}$ = (1.65$\pm$0.06)$\times$10$^{-4}$, with only a handful of outliers in the Hyades deviating from the narrow distribution around this \LLH\ level (see top panels, \autoref{fig_rossbyHa}). This value of (\LLH)$_\mathrm{sat}$ is consistent with what \citet{Newton2017} found for their sample of saturated field M dwarfs, for which \LLH\ = (1.49$\pm$0.08)$\times$10$^{-4}$, within 2$\sigma$ of our result.

On the other hand, in \citetalias{Douglas2014}, we found that, for single members in both clusters, (\LLH)$_\mathrm{sat}$ = (1.26$\pm$0.04)$\times$10$^{-4}$. Similarly, \citet{Nunez2017} found (\LLH)$_\mathrm{sat}$ = (1.27$\pm$0.02)$\times$10$^{-4}$ for single members of the $\approx$500 Myr-old cluster M37. Both of these values are statistically discrepant with our new result at the $\approx$4$\sigma$ level. However, in neither of these studies were the EW measurements corrected to account for the quiescent photospheric \halpha\ absorption. 
%, as we have (see \autoref{sec_relativeEW}). 
In addition, our new $\chi$ values used to calculate \LLH\ are $\approx$1.3$\times$ larger than those used in the two studies (see \autoref{llh}).

Accounting for quiescent absorption and using updated larger $\chi$ values results in slightly enhanced \LLH\ values, which explains our larger best-fit value for (\LLH)$_\mathrm{sat}$ compared to \citetalias{Douglas2014} and \citet{Nunez2017}.  In \autoref{app:noquiescent}, we repeated our fitting to the \Ro--\LLH\ data when the EW data are not corrected, which provides a clearer comparison to previous studies that did not apply any correction to the EW values.

Binaries and stars with RUWE $>$ 1.4 (bottom panels, \autoref{fig_rossbyHa}) exhibit a spread around the saturated level similar to that observed in single cluster stars. Their (\LLH)$_\mathrm{sat}$ value, (1.76$\pm$0.09)$\times$10$^{-4}$, is within 1$\sigma$ of that of their single counterparts.

\paragraph{The Rossby Threshold Between Saturated and Unsaturated Regimes}

For single HyPra stars, the transition between the saturated and unsaturated regimes occurs at \Ro$_\mathrm{,sat}$~=~0.29$\pm0.01$. We note that the quoted 1$\sigma$ uncertainties for \Ro$_\mathrm{,sat}$ in all of the studies under consideration, including this one, are unrealistically small, as \Ro\ uncertainties are difficult to estimate and therefore not included when running the MCMC fit. As such, we do not expect our result to statistically agree with results in similar studies. Indeed, \citet{Newton2017} found \Ro$_\mathrm{,sat}$~=~0.21$\pm0.02$, \citetalias{Douglas2014}, 0.11$^{+0.02}_{-0.03}$, and \citet{Nunez2017}, 0.03$\pm0.01$. All of these results are statistically discrepant by $\ge$3$\sigma$. 

As we show in \autoref{app:noquiescent}, however, re-running our MCMC fit without applying the quiescent absorption correction to our measured EW values results in \Ro$_\mathrm{,sat}$ values that do agree statistically with that from \citetalias{Douglas2014}, but are still statistically discrepant from that in \citet{Nunez2017}. 

Lastly, for known and candidate binaries, \Ro$_\mathrm{,sat} = 0.20^{+0.03}_{-0.04}$, which is within 2$\sigma$ of the value of single members, notwithstanding the unaccounted for uncertainties in \Ro\ mentioned above.

\begin{figure*}[t]
\centerline{\includegraphics{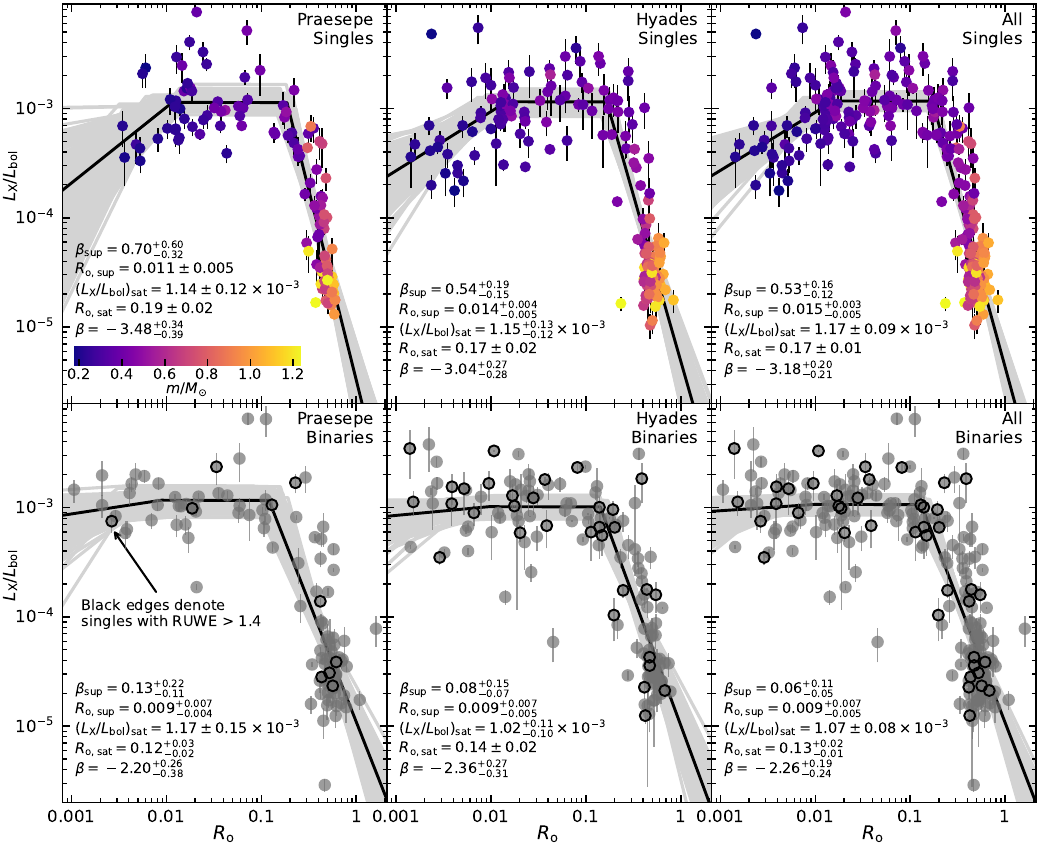}}
\caption{Same as \autoref{fig_rossbyHa}, but for \LLX. In addition to the saturated and unsaturated regimes in \autoref{fig_rossbyHa}, we also assumed the existence of a supersaturated regime described by a power-law of index $\beta_\mathrm{sup}$ and \Ro$_\mathrm{,sup}$. The result of the fit for the five parameters $\beta_\mathrm{sup}$, \Ro$_\mathrm{,sup}$, (\LLX)$_\mathrm{sat}$, \Ro$_\mathrm{,sat}$, and $\beta$, 
are given in each panel. We show in \autoref{app:dists} the marginalized posterior probability distributions from the MCMC analysis for each fit.}
\label{fig_rossbyX}
\end{figure*}

\paragraph{The Unsaturated Regime}

For single HyPra stars, we found that $\beta = -5.85^{+0.81}_{-0.80}$. This result is $>$4$\sigma$ away from that of \citet{Newton2017}, who found $\beta = -1.7$$\pm$$0.1$. Although our methods are similar to those used by these authors, our unsaturated stars are significantly different from those in \citet{Newton2017} in three ways. 

First, the majority of stars with \Ro\ $>$ \Ro$_\mathrm{,sat}$ in our sample have masses $\gtrsim$0.5 \Msun\ (see the colormap in \autoref{fig_rossbyHa}), whereas their sample does not have any stars with masses $\gtrsim$0.5 \Msun\ (see their Figure 6). Second, our largest \Ro\ values are $\approx$ 0.5, whereas most unsaturated stars in their sample have \Ro\ $>$ 0.5 and up to 2.0. And third, all of our stars are $\approx$700 Myr old, whereas their sample mostly included field-age dwarfs, which presumably have ages $>>$1 Gyr. The $\beta$ discrepancy between these two samples may partly be evidence for a steeper decay in chromospheric activity for the more massive, partly convective dwarfs versus for fully or almost fully convective dwarfs. On the other hand, the $\beta$ discrepancy may just reflect different dominant chromospheric radiative coolants for stars at different \teff: in M dwarfs, Balmer lines emission dominates, whereas in G and K dwarfs, Ca {\sc ii} and Mg {\sc ii} emission dominates \citep{Linsky1982, Reid2005}.

In \citetalias{Douglas2014}, we found that $\beta=-0.73^{+0.16}_{-0.12}$, and in \citet{Nunez2017}, $\beta=-0.51$$\pm$$0.02$. Both of these results are also statistically inconsistent with our new result. However, as noted earlier, these two studies must be compared to our results when we do not apply the quiescent correction to our EWs (see \autoref{app:noquiescent}).

For binaries and candidate binaries, we found that $\beta=-2.05$$\pm$$0.50$, which is within 3$\sigma$ of our result for single stars. The shallower $\beta$ for binaries partly reflects the slightly higher---although statistically insignificant---\halpha\ emission in binaries compared to single stars in the \gminusk\ = 3.0--3.3 mag bin ($\approx$M0--M2 stars; see \autoref{fig_ewbinned}). 

However, as we mentioned earlier, using $M_G$ to derive \teff\ and $m$, from which we then calculated $\chi$ and $\tau$, leads to overestimated \LLH\ and \Ro\ values to varying degrees for binaries. Therefore, we do not consider our shallower $\beta$ result to be evidence for higher chromospheric activity in unsaturated binaries compared to their single counterparts.

\subsection{The Dependence of Coronal Activity on Rotation} \label{sec_xrossby}

In \citetalias{Nunez2022}, we presented a comprehensive study of \LLX\ as a coronal activity indicator and of its dependence on \Ro\ in Praesepe and the Hyades. We used a sample of 114 Praesepe and 63 Hyades single stars to characterize the saturated and unsaturated regimes in the \Ro--\LLX\ plane, using the same parametrization given in Equation~\ref{eq:rossby} (we also had 107 Praesepe and 98 Hyades binary stars or with RUWE $>$ 1.4).

In that study, we found weak evidence for supersaturation (see appendix B of \citetalias{Nunez2022}), the \Ro\ regime in which super-fast rotators (\Ro\ $\lesssim$ 0.01) show a decrease in activity level relative to their saturated cousins. To characterize this behavior, we modified the \Ro--\LLX\ relation parametrization presented in Equation \ref{eq:rossby} by adding a secondary power-law at small \Ro: below $R_\mathrm{o,sup}$, activity declines as a power-law with index $\beta_\mathrm{sup}$.

Since that study, we have added 154 stars to our sample of cluster stars with both \Ro\ and \LLX\ measurements. These are primarily Hyads; we have an additional 99 single stars and 40 known and candidate binaries in that cluster with these measurements (the numbers for Praesepe are 10 and five, respectively). \autoref{fig_rossbyX} shows the updated \Ro--\LLX\ relation for single stars (top panels) and binary stars (bottom panels) for Praesepe (left panels), Hyades (middle panels), and both clusters combined (right panels). 

With this update, we found more compelling evidence of supersaturation in single stars in both clusters. In this regime, single stars in Praesepe follow a power-law of slope $\beta_\mathrm{sup} = 0.70^{+0.60}_{-0.32}$, 2$\sigma$ away from a flat relation, while in Hyades, where the number of supersaturated stars is larger, $\beta_\mathrm{sup} = 0.54^{+0.19}_{-0.15}$, 3$\sigma$ away from being flat. Meanwhile, for the combined HyPra sample 
$\beta_\mathrm{sup} = 0.53^{+0.16}_{-0.12}$, which is at least 4$\sigma$ away from a flat relation (top row, \autoref{fig_rossbyX}).

On the other hand, for known and candidate binaries, the supersaturated regime is almost indistinguishable from a flat relation ($\beta_\mathrm{sup}$ = 0) and \Ro$_\mathrm{,sup}$ remains poorly constrained (bottom panels, \autoref{fig_rossbyX}). The lack of supersaturation in binaries may be partly explained by magnetic interactions between the binary components increasing their quiescent activity levels and/or increasing the frequency of flaring activity. The updated results for the other four parameters, namely, \Ro$_\mathrm{,sup}$, (\LLX)$_\mathrm{sat}$, \Ro$_\mathrm{,sat}$, and $\beta$, only change marginally compared to our results in \citetalias{Nunez2022}, with $\beta$ being now more constrained for both single and binary members. We include these updated parameters in \autoref{tbl_rossbies}.

\subsection{Chromospheric Versus Coronal Activity} \label{compare_activity}

\begin{figure}[t]
\centerline{\includegraphics{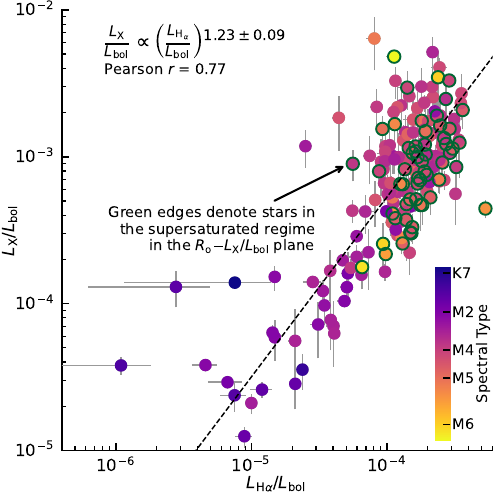}}
\caption{\LLX\ vs.~\LLH\ for single stars in Praesepe and the Hyades, color-coded by their spectral type. The dashed black line is the power-law relation found with a least-squares bisector regression. The slope and Pearson $r$ of the regression is noted at the top left. Green edges highlight stars in the supersaturated regime in the \Ro--\LLX\ plane.}
\label{fig_compHaX}
\end{figure}

Several studies have shown differences in the dependence of \halpha\ and X-ray emission on rotation \citep[e.g.,][]{Hodgkin1995,Nunez2017,Preibisch2005a,Stelzer2013}. These differences could point to differences in magnetic heating mechanisms acting on different layers of the stellar atmospheres. At the same time, some positive correlation between \LLX\ and \LLH\ is expected, partly because a fraction of the coronal X-rays will inevitably heat the underlying chromosphere \citep{Mullan1976, Cram1982}. We directly compare \LLX\ and \LLH\ for single stars in both clusters in \autoref{fig_compHaX} to characterize their relationship. 

Using a least-squares bisector regression, we found a power-law relation such that \LLX\ $\propto$ (\LLH)$^\alpha$, with $\alpha$ = 1.23$\pm0.09$ (dashed line, \autoref{fig_compHaX}), with a correlation coefficient $r$ = 0.77, which suggests a strong positive correlation between \LLX\ and \LLH.

Most of the stars are concentrated near \LLH~$\approx~10^{-4}$ and \LLX~$\approx 10^{-3}$. These two values correspond to the saturation levels in both activity indicators. The tail-like structure that goes from this locus to smaller values in both \LLH\ and \LLX\ corresponds to stars in the unsaturated regime of both indicators. Finally, we highlight in \autoref{fig_compHaX} stars in the supersaturated regime in the \Ro-\LLX\ plane, most of which lie below the power-law relation.

In \citet{Nunez2017}, we found for single members of M37 a weaker correlation ($r$ = 0.63) and a slope closer to 1:1 ($\alpha$ = 1.05$\pm$0.01). In that $\approx$500-Myr-old cluster, our sample included stars in the spectral range K0--M1 that were almost all saturated in both \LLH\ and \LLX. By contrast, our Praesepe and Hyades sample includes K6--M6 stars with \LLH\ and \LLX\ measurements (see the colorbar in \autoref{fig_compHaX}), and a significant number of these are unsaturated. In addition, the \LLH\ values in \citet{Nunez2017} did not account for the quiescent correction described in \autoref{sec_relativeEW}, the effect of which is difficult to quantify in this analysis.

By contrast, \citet{He2019} found $\alpha$~=~1.12$\pm$0.30 for a sample of field-age K and M dwarfs. This result agrees with ours at the 1$\sigma$ level. Also, for a sample of M dwarfs within 10 pc, \citet{Stelzer2013} found $\alpha$~=~1.90$\pm$0.31, implying a steeper slope for the unsaturated rotation--activity relation---but this value is in 2$\sigma$ agreement with our value for $\alpha$. The former study accounted for quiescent \halpha\ absorption, while the latter did not.

\begin{figure}[t]
\centerline{\includegraphics{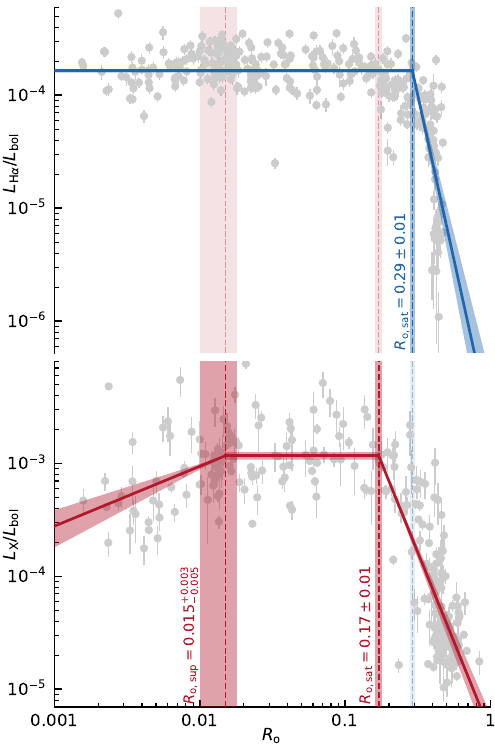}}
\caption{\LLH\ vs.~\Ro\ (top panel) and \LLX\ vs.~\Ro\ (bottom panel) for single members of Praesepe and the Hyades combined (gray circles). The maximum \textit{a posteriori} fits from the MCMC algorithm (see Sections~\ref{sec_rossbyresults} and \ref{sec_xrossby}) and their approximate 1$\sigma$ uncertainties are indicated with solid lines and shaded regions, respectively. The Rossby threshold values (\Ro$_\mathrm{,sup}$ and \Ro$_\mathrm{,sat}$) and their 1$\sigma$ uncertainties are indicated with vertical dashed lines and shaded regions, respectively, and are annotated next to each line. We extend these vertical dashed lines along both panels to more easily compare the different regimes (supersaturated, saturated, and unsaturated) in both chromospheric (\LLH) and coronal (\LLX) activity indicators.}
\label{fig_rossbyXHa}
\end{figure}

It is more informative to directly compare relations for \LLH\ and \LLX\ as a function of \Ro. We re-create the top right panel of Figures \ref{fig_rossbyHa} and \ref{fig_rossbyX}, i.e., the HyPra sample, as the top and bottom panels (respectively) in \autoref{fig_rossbyXHa}. We highlight the results from the MCMC algorithm with solid lines and shaded regions, corresponding to the maximum \textit{a posteriori} and 1$\sigma$ MCMC results. We also highlight with vertical dashed lines the threshold Rossby values, namely, \Ro$_\mathrm{,sat}$ for the \Ro--\LLH\ relation and \Ro$_\mathrm{,sup}$ and \Ro$_\mathrm{,sat}$ for the \Ro--\LLX\ relation.

In the top panel of \autoref{fig_rossbyXHa} the lack of supersaturation in \LLH\ is evident. If the fastest spinners (\Ro~$\lesssim$~0.01) appear supersaturated in X-rays but not in \halpha, then whatever mechanism is curtailing the magnetically driven X-ray emission is present in the coronae of these stars, but not in their chromospheres. 

Of the two most invoked mechanisms to explain supersaturation, centrifugal stripping of the corona \citep{Jardine1999} and reduction of the filling factor \citep{Stepien2001}, our evidence favors the former, echoing the conclusions of, e.g., \citet{Marsden2009, Jackson2010, Wright2011}. In the centrifugal stripping scenario, the chromospheric layers would not be affected by the stars' super rapid rotation, whereas in the reduced filling factor scenario, all atmospheric layers would be impacted. The centrifugal stripping scenario would not conflict with the expectation that some of the chromospheric heating comes from X-rays emitted in the corona. It is reasonable to expect that most of the X-rays heating the chromosphere originate in the denser inner layers of the corona. Thus, it is possible for \LLX\ to decrease due to plasma loss at the outermost layers of the corona, while maintaining \LLH\ mostly unaffected.

\begin{figure}
\centerline{\includegraphics{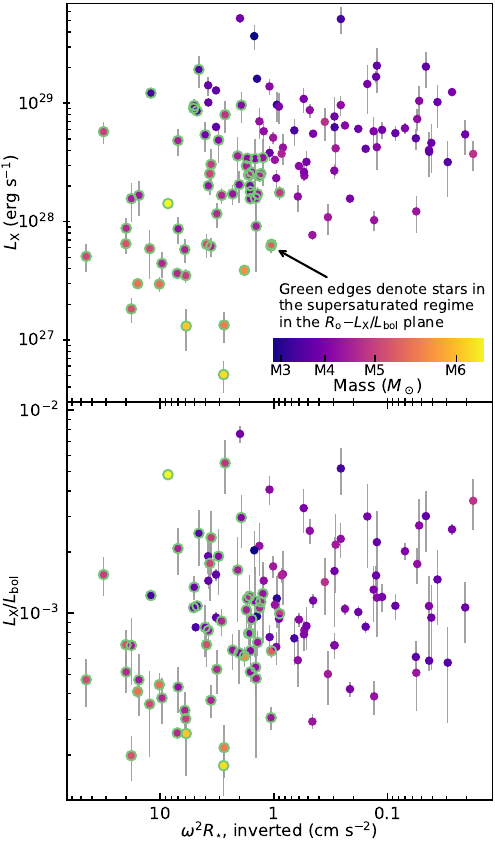}}
\caption{\LX\ (top panel) and \LLX\ (bottom panel) vs. centrifugal acceleration for single members of Praesepe and the Hyades that are in the saturated or supersaturated regimes in the top right panel of Figure \ref{fig_rossbyX}. Stars are color coded by their spectral type. Green edges highlight stars in the supersaturated regime in the \Ro--\LLX\ plane. The x axis is inverted (i.e., acceleration increases toward the left) for easier comparison to Figure \ref{fig_rossbyX}.}
\label{fig_acc}
\end{figure}

As an additional test of whether we are seeing evidence of centrifugal stripping, we compared X-ray activity to the stellar centrifugal acceleration, defined as the square of the angular rotation frequency (i.e., the reciprocal of \prot) times the stellar radius: $\omega^2R_\mathrm{\star}$. We derived $R_\mathrm{\star}$ and 1$\sigma$ uncertainties for main sequence cluster stars using the empirical $R_\mathrm{\star}$--$M_G$ relation of E.~Mamajek, and they are included in Table \ref{tbl_measurements}. In Figure \ref{fig_acc} we plot \LX\ and \LLX\ vs. centrifugal acceleration for Praesepe and Hyades single members with \Ro\ $<$ \Ro$_\mathrm{,sat}$, and we highlight with green circles those stars with \Ro\ $<$ \Ro$_\mathrm{,sup}$. We find all stars in the supersaturated regime to have $\omega^2R_\mathrm{\star}$ \gapprox\ 1 cm s$^{-2}$. We see an indication of supersaturation at $\omega^2R_\mathrm{\star}$ \gapprox\ 3 cm s$^{-2}$, although not as clear as in the \LLX--\Ro\ plane.

Also evident in \autoref{fig_rossbyXHa} is the smaller \Ro$_\mathrm{,sat}$ for \LLX\ compared to \LLH---6$\sigma$ away from each other.\footnote{As discussed in \autoref{sec_rossbyresults}, our \Ro$_\mathrm{,sat}$ uncertainties are likely underestimated. Therefore, the difference between the two \Ro$_\mathrm{,sat}$ values may not be as pronounced.} This difference indicates that the transition from the saturated to unsaturated regimes does not occur in tandem in these two layers of the stellar atmosphere. Our HyPra sample suggests that as stars spin down (i.e., their \Ro\ increases), saturation ends in the corona before it ends in the chromosphere.\footnote{ In \citet{Nunez2017}, we also found a difference in the two \Ro$_\mathrm{,sat}$ values for our sample of M37 stars, but the result was the opposite: \Ro$_\mathrm{,sat}$ was smaller for \LLH\ than for \LLX. However, as we describe in \autoref{app:noquiescent}, the M37 sample was significantly smaller and our \halpha\ measurements were contaminated by emission from a foreground nebula, both of which undermined our analysis.} This difference in timing could indicate a difference in the sensitivity to field components of the magnetic field at different atmospheric altitudes, which would not be surprising. For example, \citet{See2019} found that \Ro$_\mathrm{,sat}$ for an activity indicator derived from Zeeman-Doppler imaging, which is particularly sensitive to large-scale components of the magnetic field \citep[e.g.][]{Brown1991}, is smaller than that of other activity indicators. Based on our results, we infer that \LLX\ is more sensitive to large-scale components than \LLH.

%%%%%%%%%%%%%%%%%%%%%%%%%%%%%%%%%%%%%%%%%%%%%%%%%%%%%%%%%%%%%%
\section{Conclusion}\label{sec_conclusions}

We have performed an analysis of chromospheric and coronal activity in low-mass stars in the Praesepe and Hyades open clusters. These two coeval groups of stars, with a crucial age between that of very young clusters and that of field stars, are pivotal in our understanding of the dependence of stellar magnetic activity on rotation and of the evolution of this dependence. 

We used the Praesepe and Hyades membership catalogs of \citetalias{Nunez2022}, which include several stellar parameters such as mass, distance, \Lbol, \prot, $\tau$, and binarity identification, as well as Gaia and 2MASS photometry. We updated these quantities when appropriate (e.g., to Gaia DR3 values), and added to the catalogs the ratio of the continuum flux near the \halpha\ line to the apparent bolometric flux, $\chi$, and \teff\ for most stars.

We gathered several hundred new optical spectra using the MDM and MMT Observatories to complement our sample of existing spectra, published nearly a decade ago in  \citetalias{Douglas2014}. We complemented these new spectra with spectra from the public SDSS and LAMOST catalogs. For a few hundred cluster stars we have multiple high-quality spectra. We also obtained new X-ray detections and \LX\ measurements for an additional ten Praesepe stars and 23 Hyads.

To complement the existing rotational data for Praesepe and Hyades stars, we measured \prot\ values using TESS and ZTF light curves for an additional 28 Praesepe stars and 137 Hyads.

From our optical spectra, we measured the \halpha\ EW and then estimated a relative EW value after accounting for the quiescent photospheric \halpha\ absorption present in low-mass stars. We then estimated \LLH\ by using our relative EW values and an expanded version of our previously published $\chi$--\teff\ relation based on PHOENIX model spectra. In the color--EW plane, we find that at $\approx$700 Myr all late F, G, and K type dwarfs have converged onto a tight sequence of \halpha\ absorption, and that by contrast nearly all M dwarfs exhibit some level of \halpha\ emission. In both clusters, the transition between \halpha\ absorption and emission occurs at the same spectral type, approximately M0--M1. We also find that binaries follow the same EW distribution as their single counterparts, suggesting negligible enhancement of chromospheric activity in binary systems in the two clusters.

In the \Ro--\LLH\ plane for the combined sample of single stars from both clusters, we found a saturated regime for stars with \Ro\ $\lesssim$ 0.3, with a saturation level (\LLH)$_\mathrm{sat} \approx 1.7 \times 10^{-4}$. We found an unsaturated regime described by a power-law with slope $\beta \approx -5.8$ for single members and $\approx$$-2.0$ for binaries; the former is significantly steeper than the slopes found in similar studies in the literature. This difference may partly be explained by the quiescent photospheric correction we implemented and by the updated $\chi$ values we used. Nonetheless, our unsaturated stars include many more massive stars ($\gtrsim$0.5 \Msun) than samples in the literature, which may be driving the steepness of the power-law fit. This steeper slope may be evidence for a more rapid decay in chromospheric activity for partly convective stars compared to their fully or almost fully convective counterparts. Alternatively, the steeper slope may just reflect a shift in chromospheric radiative cooling mechanism from Balmer lines in the cooler M dwarfs to Ca {\sc ii} and Mg {\sc ii} lines in the hotter G and K dwarfs. Finally, we found no evidence for supersaturation in \LLH.

We updated the \Ro--\LLX\ analysis in \citetalias{Nunez2022} by including our expanded sample of new stars with \prot\ and \LX\ measurements. This resulted in compelling evidence for supersaturation in \LLX\ in single stars. At \Ro\ $\lesssim$ 0.01, \LLX\ decreases following a power-law with slope $\beta_\mathrm{sup} \approx 0.5$. For binaries, on the other hand, we found no evidence for supersaturation. 

A comparison of \LLH\ and \LLX\ of Praesepe and Hyades single members revealed a close to 1:1 relation. However, stars are less well-defined by this 1:1 relation at \LLX\ $\approx 10^{-3}$ and \LLH\ $\approx 10^{-4}$, which correspond to the activity levels of saturated stars in the two activity indicators.

As Praesepe and Hyades stars show supersaturation at \Ro\ $\lesssim$ 0.01 in the coronal activity indicator (\LLX) and not in the chromospheric indicator (\LLH), our data favor centrifugal stripping as the most likely explanation for this supersaturation. Estimating the centrifugal acceleration in these stars also provides some evidence for centrifugal stripping. Also, a smaller \Ro$_\mathrm{,sat}$ for the coronal activity indicator compared to the chromospheric indicator may be evidence for a higher sensitivity of \LLX\ to large-scale magnetic field components.

%%%%%%%%%%%%%%%%%%%%%%%%%%%%%%%%%%%%%%%%%%%%%%%%%%%%%%%%%%%%%%
\section*{Acknowledgments}

We thank the anonymous referee for their helpful comments. We thank Matthew Browning for insightful discussions on stellar magnetic fields. We thank Elliana S.~Abrahams, Emily C.~Bowsher, Theron Carmichael, Victoria DiTomasso, Rose K.~Gibson, David Jaimes, Andy Lawler, Mark Popinchalk, Justin Rupert, Ian Weaver, and Jos\'e M.~Zorrilla for their assistance in obtaining spectra for this work at the MDM Observatory.

A.N.~acknowledges support provided by the NSF through grant 2138089. M.A.A.~acknowledges support from a Fulbright U.S.~Scholar grant co-funded by the Nouvelle-Aquitaine Regional Council and the Franco-American Fulbright Commission and from a Chr\'etien International Research Grant from the American Astronomical Society. M.A.A.~also acknowledges support from NASA TESS grant 80NSSC19K0383 (program ID G011197) and NASA grant 80NSSC21K0989. J.L.C.~is supported by NSF grant AST-2009840 and NASA TESS grant 80NSSC22K0299 (program ID G04217). K.R.C.~acknowledges support provided by the NSF through grant AST-1449476. J.J.D.~was supported by NASA contract NAS8-03060 to the Chandra X-ray Center.

This work is based on observations obtained at the MDM Observatory, operated by Dartmouth College, Columbia University, Ohio State University, Ohio University, and the University of Michigan. The authors are honored to be permitted to conduct astronomical research on Iolkam Du'ag (Kitt Peak), a mountain with particular significance to the Tohono O'odham.

This paper includes data collected by the TESS mission. Funding for the TESS mission is provided by the NASA's Science Mission Directorate.

This work is based on observations obtained with the Samuel Oschin Telescope 48-inch and the 60-inch Telescope at the Palomar Observatory as part of the Zwicky Transient Facility project \citep{Masci2019}. ZTF is supported by the National Science Foundation under Grants No. AST-1440341 and AST-2034437 and a collaboration including current partners Caltech, IPAC, the Weizmann Institute for Science, the Oskar Klein Center at Stockholm University, the University of Maryland, Deutsches Elektronen-Synchrotron and Humboldt University, the TANGO Consortium of Taiwan, the University of Wisconsin at Milwaukee, Trinity College Dublin, Lawrence Livermore National Laboratories, IN2P3, University of Warwick, Ruhr University Bochum, Northwestern University and former partners the University of Washington, Los Alamos National Laboratories, and Lawrence Berkeley National Laboratories. Operations are conducted by COO, IPAC, and UW.

Some of the data presented in this paper were obtained from the Mikulski Archive for Space Telescopes (MAST). STScI is operated by the Association of Universities for Research in Astronomy, Inc., under NASA contract NAS5-26555. Support for MAST for non-HST data is partly provided by the NASA Office of Space Science via grant NNX09AF08G.

Funding for the Sloan Digital Sky Survey IV has been provided by the Alfred P. Sloan Foundation, the U.S. Department of Energy Office of Science, and the Participating Institutions. SDSS acknowledges support and resources from the Center for High-Performance Computing at the University of Utah.

This work has benefited from the public data released from the Guoshoujing Telescope (Large Sky Area Multi-Object Fiber Spectroscopic Telescope, or LAMOST), a National Major Scientific Project built by the Chinese Academy of Sciences. Funding for the project has been provided by the National Development and Reform Commission. LAMOST is operated and managed by the National Astronomical Observatories, Chinese Academy of Sciences.

This research has made use of data obtained from the 4XMM XMM-Newton Serendipitous Source Catalog compiled by the 10 institutes of the XMM-Newton Survey Science Centre selected by ESA, and of data obtained from the Chandra Source Catalog, provided by the Chandra X-ray Center (CXC) as part of the Chandra Data Archive.

This work has made use of data from the European Space Agency (ESA) mission Gaia (\url{https://www.cosmos.esa.int/gaia}), processed by the Gaia
Data Processing and Analysis Consortium (\url{https://www.cosmos.esa.int/web/gaia/dpac/consortium}), and funded by national institutions. Gaia data was accessed via the VizieR database of astronomical catalogs \citep{Ochsenbein2000}.

This research has made use of the NASA/IPAC Infrared Science Archive, which is operated by the Jet Propulsion Laboratory, California Institute of Technology, under contract with the National Aeronautics and Space Administration. The Two Micron All Sky Survey was a joint project of the University of Massachusetts and IPAC.

This research has made use of NASA's Astrophysics Data System Bibliographic Services and the SIMBAD database, operated at CDS, Strasbourg, France.

PyRAF is a product of the Space Telescope Science Institute, which is operated by AURA for NASA. IRAF is distributed by the National Optical Astronomy Observatories, which are operated by the Association of Universities for Research in Astronomy, Inc., under cooperative agreement with the National Science Foundation.

\facilities{CDS, CXO, Gaia, Hiltner (OSMOS, Modspec), LAMOST, MMT (Hectospec), PO:1.2m (ZTF), ROSAT, Sloan, Swift, TESS, XMM}

\software{astropy \citep{astropyIII}}, emcee \citep{Foreman2013}, Matplotlib \citep{Hunter2007}, NumPy \citep{Harris2020}, PHEW \citep{PHEW}, PypeIt \citep{Prochaska2020}, PyRAF \citep{pyraf}, PySpecKit \citep{Ginsburg2011}, SciPy \citep{scipy}, tesscheck \citep{tesscheck}, tesscut \citep{tesscut}, unpopular \citep{Hattori2022}

\setlength{\baselineskip}{0.6\baselineskip}
\bibliography{main}
\bibliographystyle{aasjournal}
\setlength{\baselineskip}{1.667\baselineskip}

\clearpage

%%%%%%%%%%%%%%%%%%%%%%%%%%%%%%%%%%%%%%%%%%%%%%%%%%%%%%%%%%%%%%
\appendix

\section{Rotation Analysis with TESS and ZTF Light Curves}\label{app:lcs}

\subsection{An Example Showcasing the Differing Qualities of TESS and ZTF Data}\label{app:lcexample}

\begin{figure*}
\centerline{\includegraphics[width=\columnwidth]{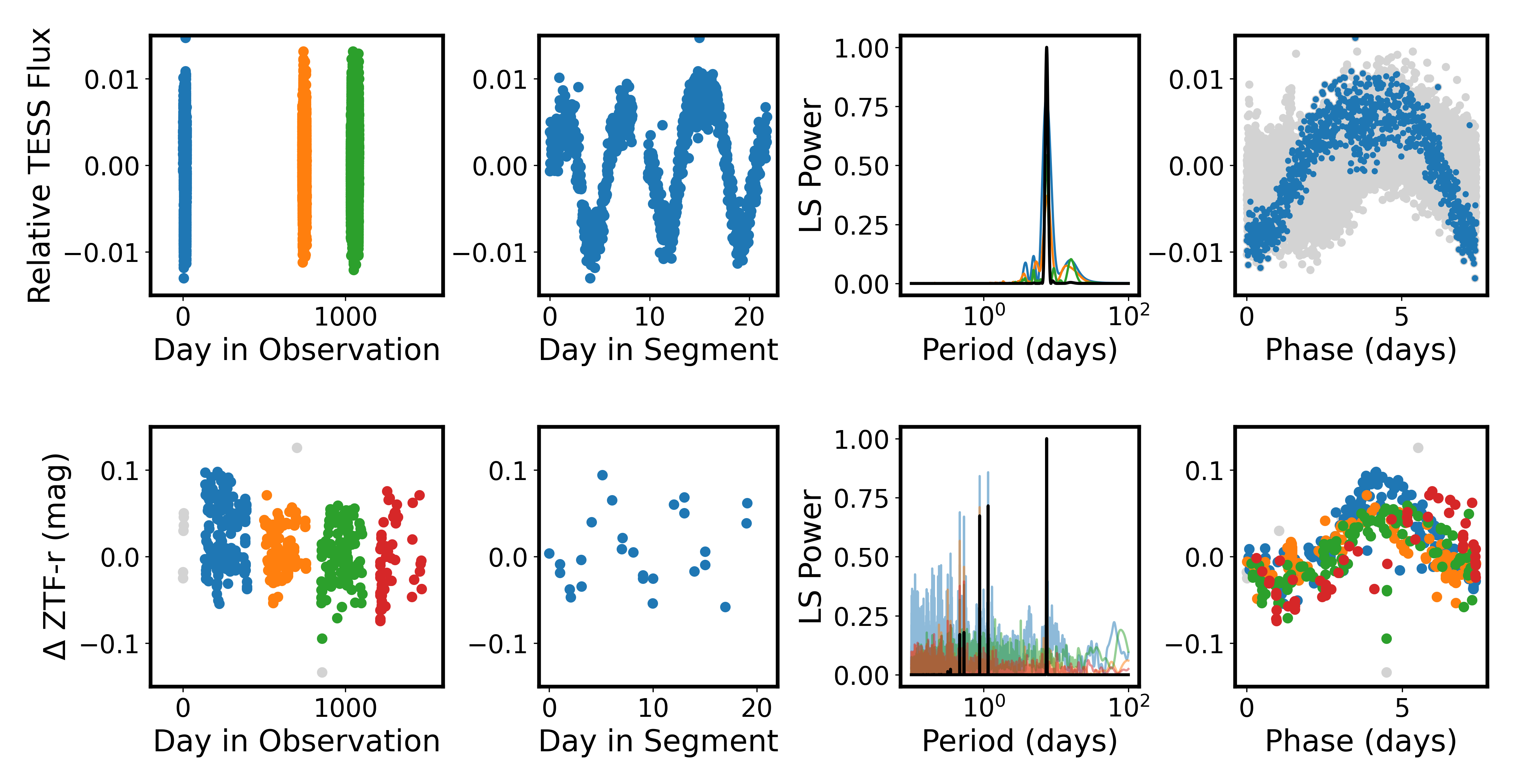}}
\caption{TESS and ZTF light curves and period analysis for the Hyad \object{2MASS J05512353+1533043} (Gaia~DR3 3348035553945613952). Top panels show TESS data and bottom panels show ZTF data. The leftmost panels show all available light curve data (at the time of our analysis), with each survey's time reference to their first epochs (i.e., both start at zero time). The center-left panels show representative segments of length equal to a single TESS sector ($\approx$27~days). The center-right panels show Lomb--Scargle periodograms for each segment (color-coded for individual sectors for TESS; different seasons for ZTF), and the full data set (black). The rightmost panels present the phase-folded light curves. Despite differences in data quality (total baseline, duration of each sector/season, cadence, precision, angular resolution), the periods precisely agree.}
\label{fig_lc_example}
\end{figure*}

Figure~\ref{fig_lc_example} presents the TESS and ZTF light curve data for the Hyad \object{2MASS J05512353+1533043} (\object{Gaia DR3 3348035553945613952}), spectral type $\approx$M3.5. At the time our analysis was performed, TESS had observed this target in Sectors 6, 33, 44, and 45 ($\Delta t \approx$  1082~days, $N =$ 10,864 observations); ZTF had observed this star over four seasons ($\Delta t \approx$ 1469~days, $N =$ 708 observations). The second column of Figure~\ref{fig_lc_example} zooms in on a representative $\approx$27-d segment (approximately the length of one TESS sector); this highlights the vast difference in cadence between ZTF (approximately nightly) and TESS (collected every 30 min during Cycle 6, and every 10 min for the later sectors operating during the first extended mission). 

The third panel shows Lomb--Scargle periodograms for the individual sectors/seasons (color-coded according to the light curves plotted in the first column) and for each full data set (black). Due to the $\sim$nightly cadence, periodograms for ZTF light curves often show high-frequency peaks near the 1-day sampling alias. Fortunately in this case, the true period has a higher power and is corroborated by the TESS periodogram, which is immune to such aliasing. The extended baseline for each ZTF season, and the consistency between seasons, ensures that the period recovered is likely the true period and not a half-period harmonic. Together, ZTF and TESS provide a powerful opportunity for deriving accurate and precise rotation periods than can be derived from either survey alone. However, as ZTF saturates at $G \approx$13~mag, and measuring periods with TESS for stars fainter than $G \gtrsim$~16 and $P_{\rm rot} >$~12~d becomes challenging, they also complement each other and enable the derivation of a more complete rotational census than can be done with either alone. 

In this example, we measured \prot\ = 7.39$\pm$0.11~d with TESS and \prot\ = 7.43$\pm$0.03~d with ZTF, where the uncertainties are the standard deviations among the Sectors/seasons. In other cases, we found evidence for longer periods (15--30~d) with TESS but could not determine the period due to the Sector duration; with ZTF, however, we were able to clearly determine the long period, while ruling out the nightly alias periods thanks to TESS.

\subsection{Two Rapidly Rotating Hyads With Discrepancies Between TESS and ZTF}\label{app:lcdisagreement}

\begin{figure*}
\centerline{\includegraphics[width=0.8\columnwidth]{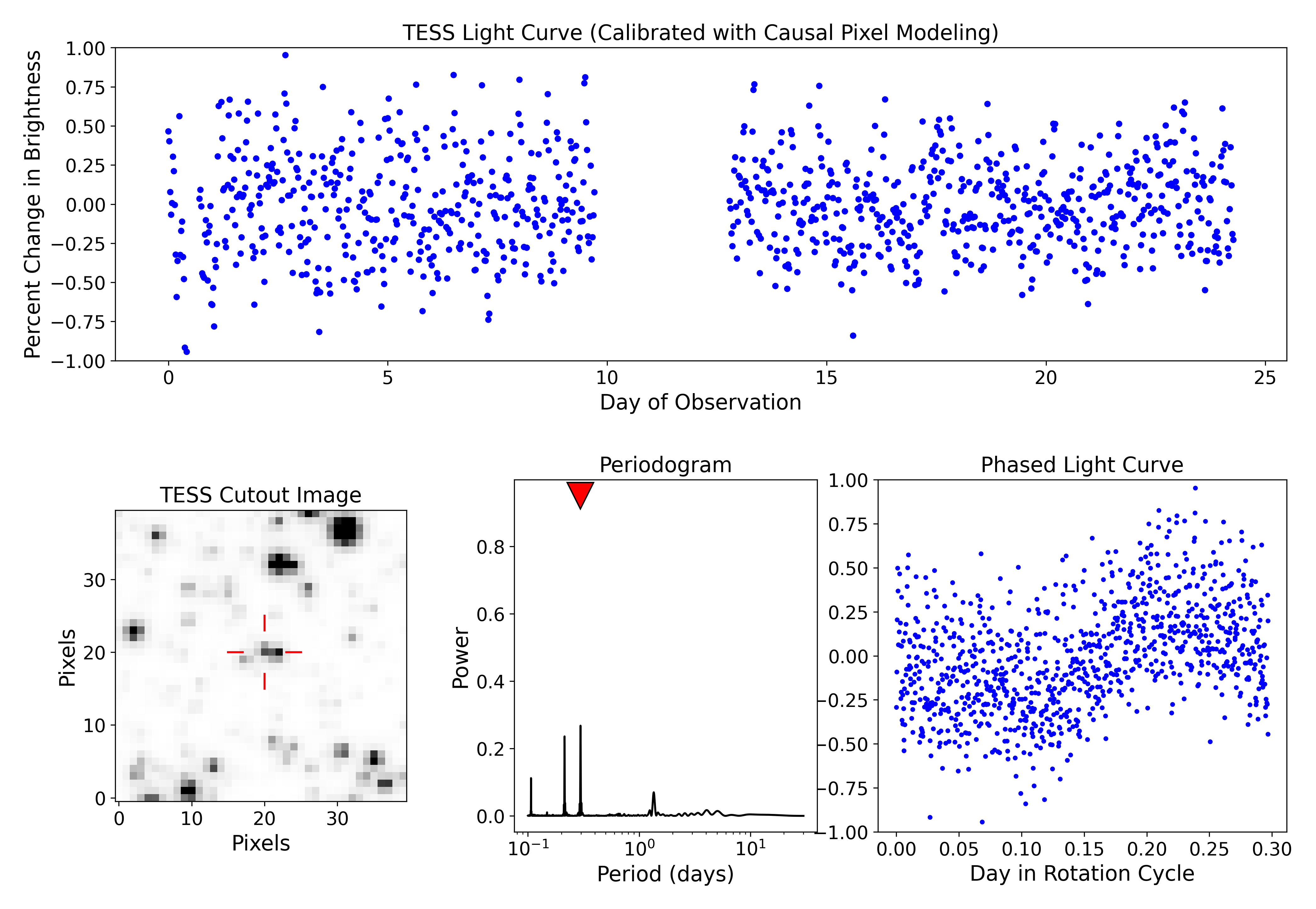}}
\centerline{\includegraphics[width=0.8\columnwidth]{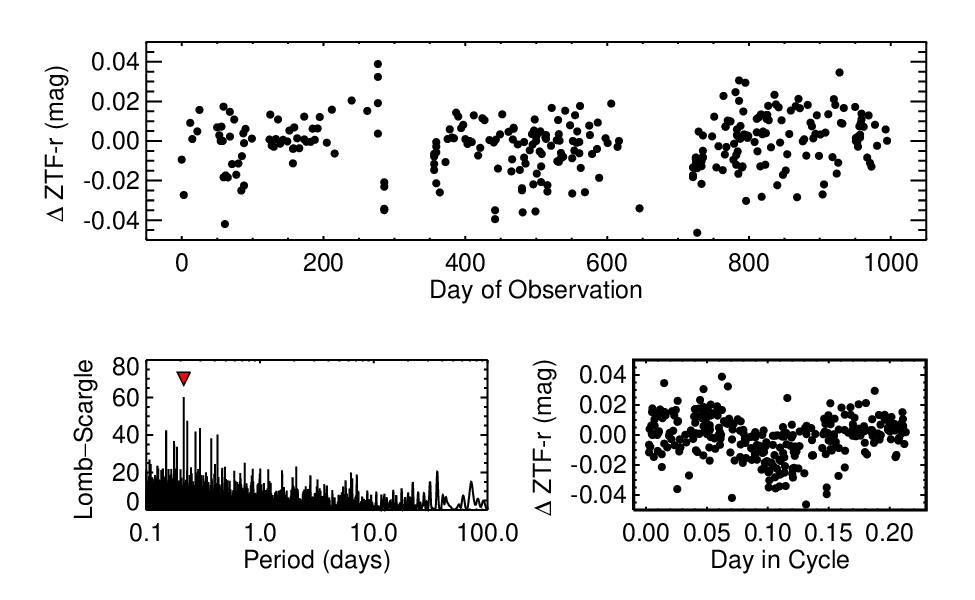}}
\caption{TESS and ZTF light curve analysis for the Hyad \object{2MASS J02594633+3855363}. Top panel: TESS light curve calibrated with Causal Pixel Modeling showing the characteristic gap midway through the 27-day observations. Middle top left: a TESS cutout indicating the position of the star. Middle top center: the Lomb-Scargle periodogram for the TESS light curve shows at least four significant peaks; the one with the highest power is indicated with a red triangle (0.30 d). Middle top right: the phase-folded TESS light curve for the 0.30 d period. Middle bottom: ZTF $r$-band light curve for three seasons. Bottom left: Lomb-Scargle periodogram for ZTF shows a weak peak at 0.21 days. Bottom right: the phase-folded ZTF light curve. We adopt the primary TESS period for this star.}
\label{fig_lcanalysis1}
\end{figure*}

\begin{figure*}
\centerline{\includegraphics[width=0.8\columnwidth]{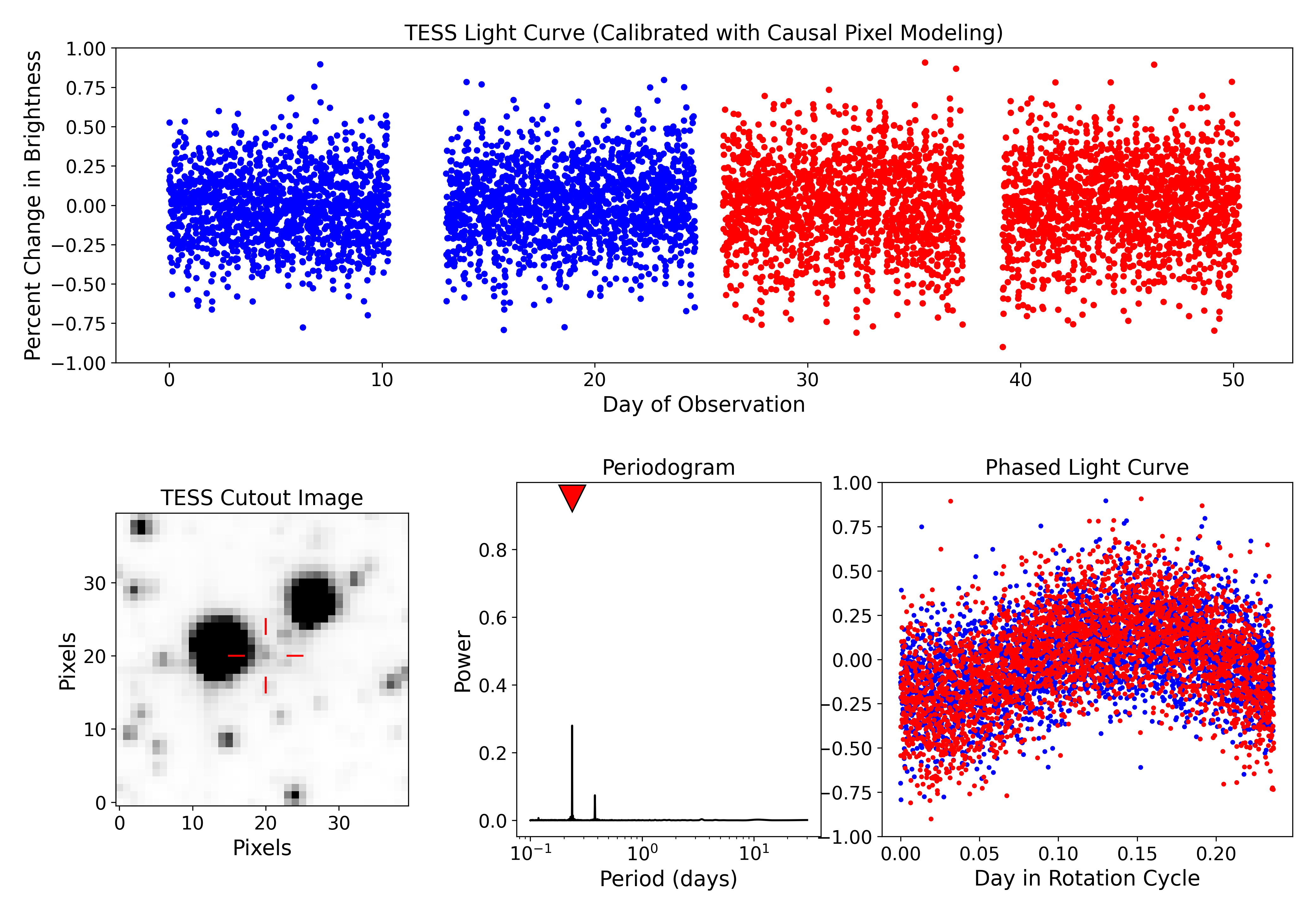}}
\centerline{\includegraphics[width=0.8\columnwidth]{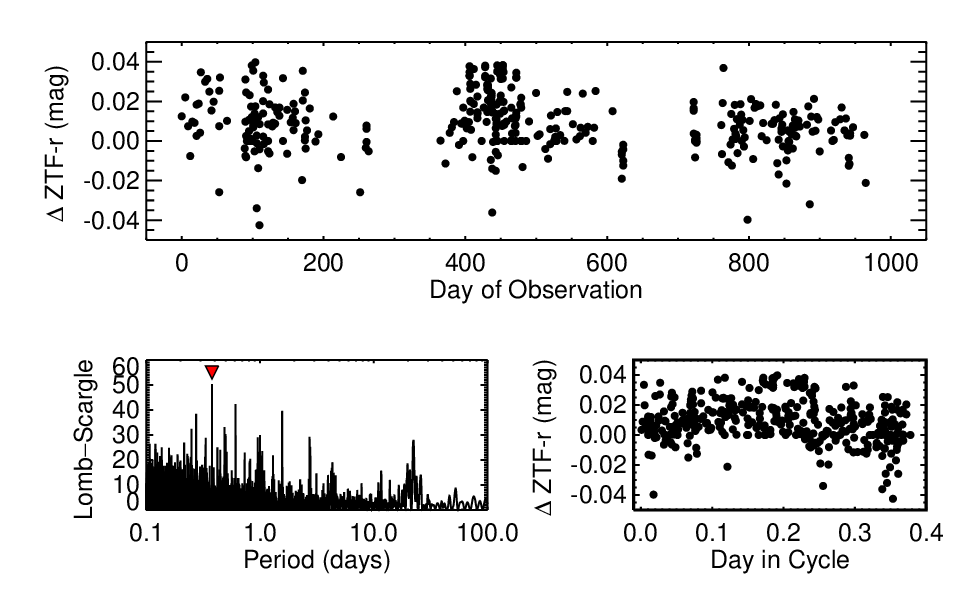}}
\caption{Same as \autoref{fig_lcanalysis1} but for the Hyad \object{2MASS J04461522+1846294}. The TESS periodogram shows two significant peaks, the one with the most power being 0.24 d. The ZTF periodogram shows a peak at 0.38 days, but presents an unconvincing phase-folded light curve, so we reject the ZTF period and adopt the primary TESS period for this star.}
\label{fig_lcanalysis2}
\end{figure*}

We present the TESS and ZTF light-curve analyses for two Hyads that have discrepant TESS and ZTF \prot\ values discussed in Section~\ref{sec_prots}: \object{2MASS J02594633+3855363} (Gaia~DR3 143558461530827264) in Figure~\ref{fig_lcanalysis1} and \object[2MASS J04461522+1846294]{J04461522+1846294} (Gaia~DR3 3409867964719693824) in Figure~\ref{fig_lcanalysis2}. 

For the first star, the TESS periodogram shows multiple rapid peaks; the most prominent has a period of 0.3~d. The ZTF periodogram shows some weak peaks in the 0.1--1.0~d range---the highest peak corresponds to 0.2~d and it looks convincingly periodic in the phase-folded light curve. This ZTF period appears to be represented in the TESS periodogram by the second highest peak. Given the high RUWE for this star (=5.0), we conclude it is likely a binary and TESS is detecting periods from both binary components. Perhaps the ZTF periodogram does not show the other significant periods because of the cadence. We adopt the primary TESS period for this star.

For the second star, the TESS periodogram shows two peaks, which are not harmonics. The primary TESS period is 0.24~days, whereas the primary ZTF period is 0.38~d. Although the ZTF periodogram and phase-folded light curves are not convincing on their own, the ZTF period is consistent with the secondary peak in the TESS light curve. For that reason, we consider the two periods detected by TESS to be hosted by the same target (and not caused by an unrelated star blended in the large TESS pixel). As in the first case, this target also boasts an elevated RUWE of 3.4, which indicates that the target is likely a binary. We assign the period for the primary peak in the TESS periodogram as the period for the primary star of the binary, although it is also possible that we have attributed the period to the wrong binary component. However, if that is the case, it will not impact the conclusions of our work: first, the Rossby number for either period places this star in the saturated regime; second, we flag all candidate and confirmed binaries and analyze them separately from the single-star cohort, the latter being the focus of our work.
%Finally, we note that there are stars which were too faint to detect periods with TESS, but have periods determined from ZTF. Given that these two discrepancies between TESS and ZTF both have elevated RUWE supporting their status as binaries, we can similarly use Gaia RUWE to flag other potentially problematic stars where we do not have usable TESS data to clearly detect periods from multiple binary component stars.

\clearpage

\section{Marginalized Posterior Probability Distributions for the MCMC Analysis of Our \Ro--\LLH\ and \Ro--\LLX\ Models}\label{app:dists}

\begin{figure}
\centerline{\includegraphics{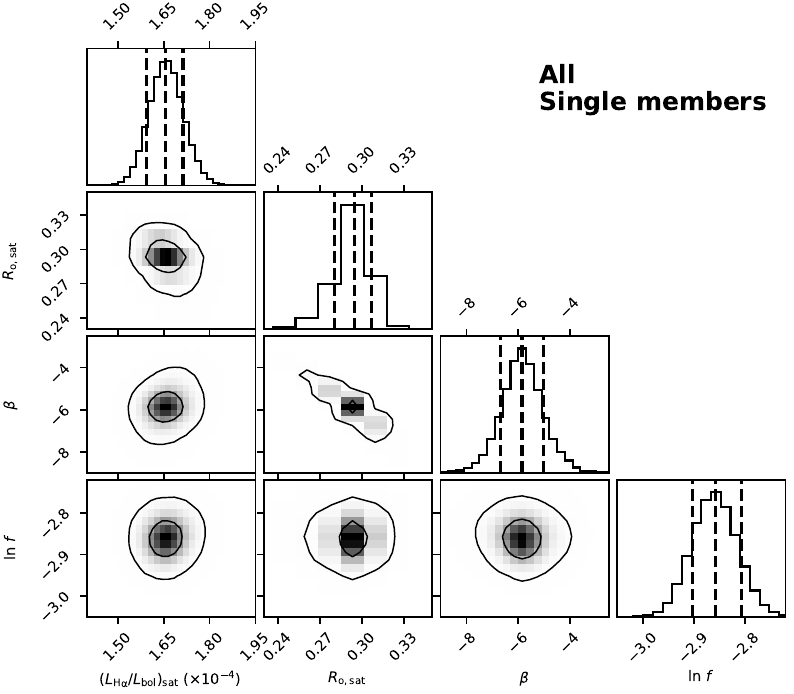}}
\caption{Marginalized posterior probability distributions from the MCMC analysis of our \Ro--\LLH\ model using \texttt{emcee} for single members in both Praesepe and the Hyades. The parameter values of the \textit{a posteriori} model are the peaks of the one-dimensional distributions; the vertical dashed lines approximate the median and 16$^{th}$, 50$^{th}$, and 84$^{th}$ percentiles. The two-dimensional distributions illustrate covariances between parameters; the contour lines approximate the 1$\sigma$ and 2$\sigma$ levels of the distributions. The complete figure set, which includes an image for each of the six subsamples in \autoref{tbl_rossbies} and for both \Ro--\LLH\ and \Ro--\LLX, is available in the online journal.}
\label{fig_posteriors}
\end{figure}

We present the marginalized posterior probability distributions from the MCMC analysis we performed on six different subsamples of Praesepe and Hyades stars: single members of each cluster, binary members of each cluster, single members of both clusters combined, and binary members of both clusters combined (see \autoref{sec_rossbyresults} and \autoref{tbl_rossbies}). The binary samples include candidate and confirmed binaries, which includes stars with RUWE $>$ 1.4. \autoref{fig_posteriors} shows an example of the marginalized posterior probability distributions for the combined sample of single members from both clusters for the \Ro--\LLH\ model.

\figsetstart
\figsetnum{15}
\figsettitle{Marginalized posterior probability distributions from the MCMC analysis of Our \Ro--\LLH\ and \Ro--\LLX\ Models}

\figsetgrpstart
\figsetgrpnum{15.1}
\figsetgrptitle{Single members of Praesepe}
\figsetplot{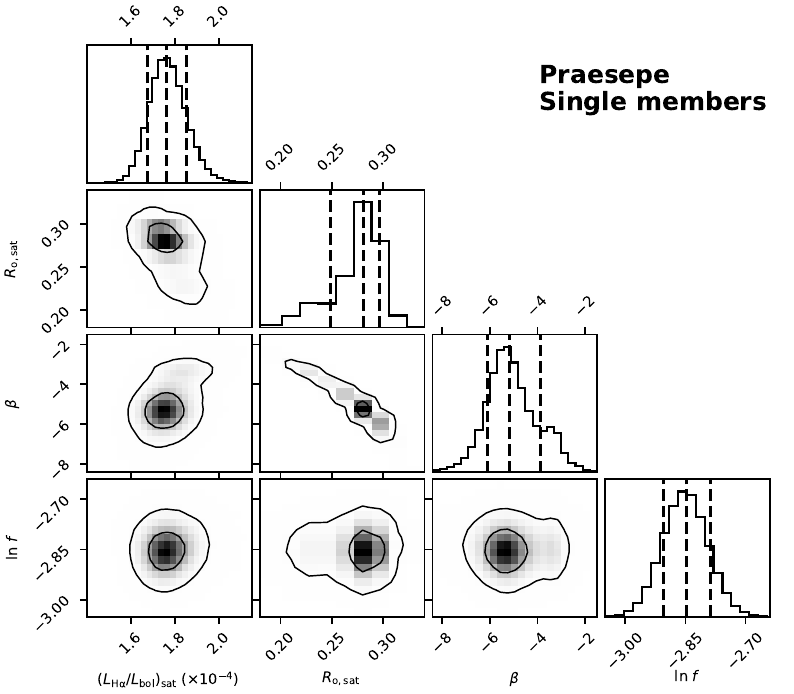}
\figsetgrpnote{Marginalized posterior probability distributions from the MCMC analysis of our \Ro--\LLH\ model using \texttt{emcee} for single members of Praesepe. The parameter values of the a posteriori model are the peaks of the one-dimensional distributions; the vertical dashed lines approximate the median and 16$^{th}$, 50$^{th}$, and 84$^{th}$ percentiles. The two-dimensional distributions illustrate covariances between parameters; the contour lines approximate the 1$\sigma$ and 2$\sigma$ levels of the distributions.}
\figsetgrpend

\figsetgrpstart
\figsetgrpnum{15.2}
\figsetgrptitle{Single members of Hyades}
\figsetplot{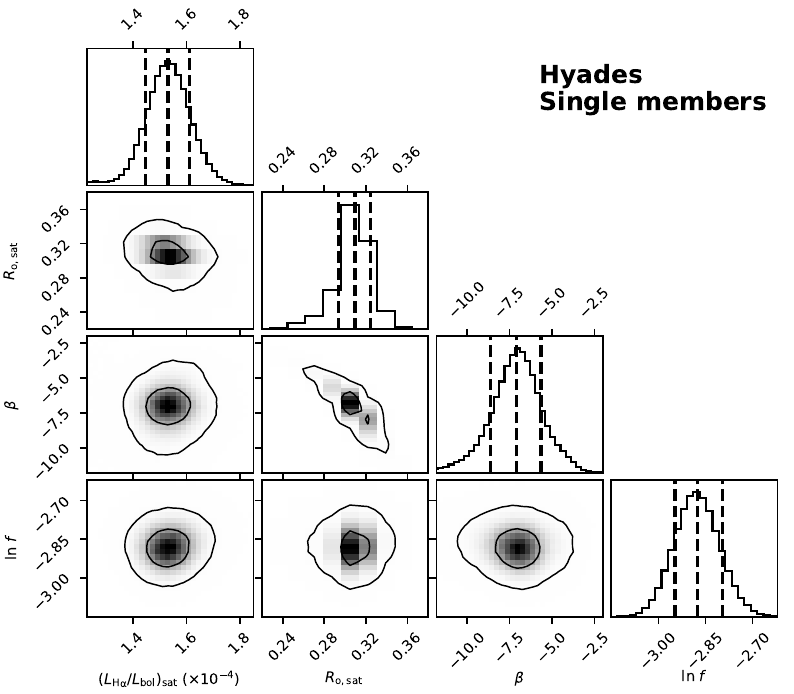}
\figsetgrpnote{Marginalized posterior probability distributions from the MCMC analysis of our \Ro--\LLH\ model using \texttt{emcee} for single members of Hyades. The parameter values of the a posteriori model are the peaks of the one-dimensional distributions; the vertical dashed lines approximate the median and 16$^{th}$, 50$^{th}$, and 84$^{th}$ percentiles. The two-dimensional distributions illustrate covariances between parameters; the contour lines approximate the 1$\sigma$ and 2$\sigma$ levels of the distributions.}
\figsetgrpend

\figsetgrpstart
\figsetgrpnum{15.3}
\figsetgrptitle{Single members of both clusters}
\figsetplot{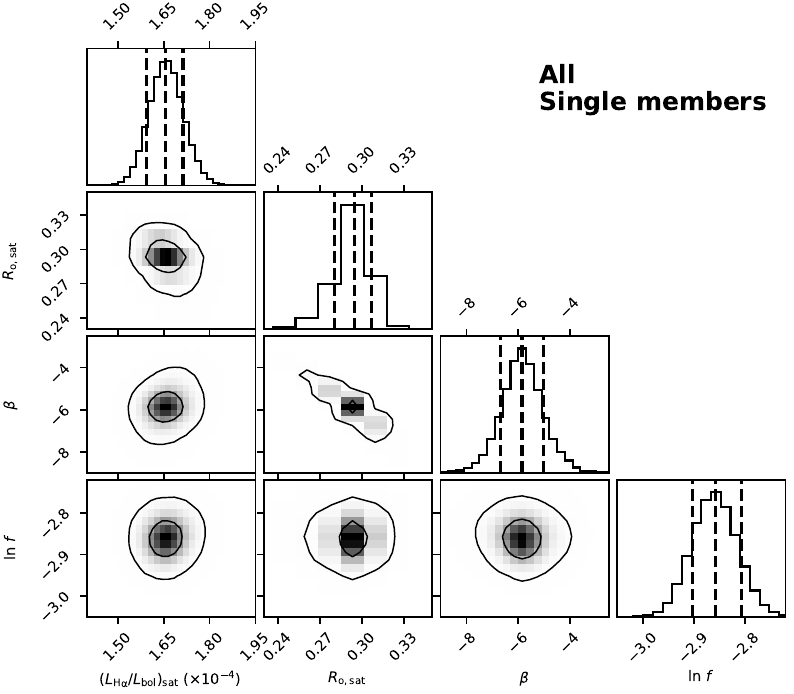}
\figsetgrpnote{Marginalized posterior probability distributions from the MCMC analysis of our \Ro--\LLH\ model using \texttt{emcee} for single members of Praesepe and Hyades together. The parameter values of the a posteriori model are the peaks of the one-dimensional distributions; the vertical dashed lines approximate the median and 16$^{th}$, 50$^{th}$, and 84$^{th}$ percentiles. The two-dimensional distributions illustrate covariances between parameters; the contour lines approximate the 1$\sigma$ and 2$\sigma$ levels of the distributions.}
\figsetgrpend

\figsetgrpstart
\figsetgrpnum{15.4}
\figsetgrptitle{Binary members of Praesepe}
\figsetplot{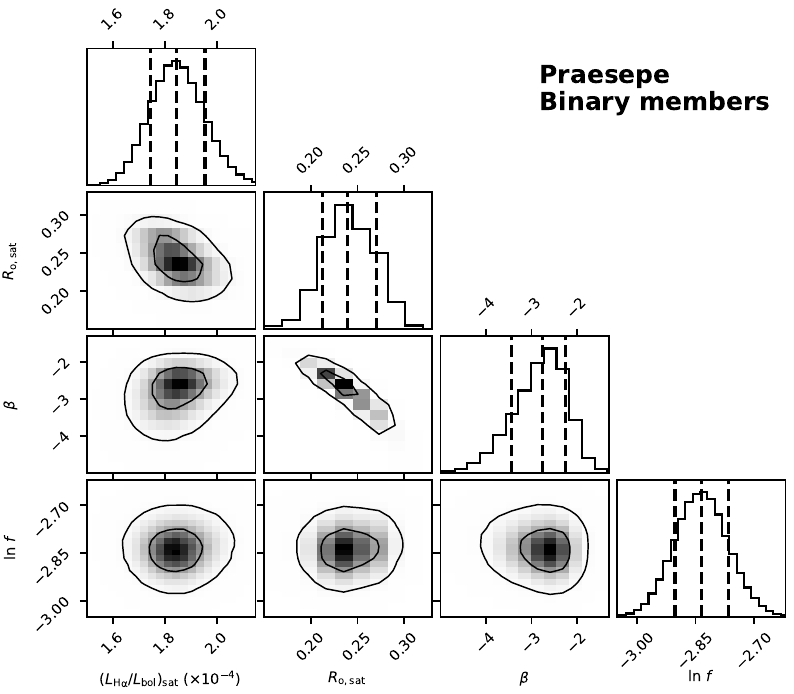}
\figsetgrpnote{Marginalized posterior probability distributions from the MCMC analysis of our \Ro--\LLH\ model using \texttt{emcee} for binary members of Praesepe. The parameter values of the a posteriori model are the peaks of the one-dimensional distributions; the vertical dashed lines approximate the median and 16$^{th}$, 50$^{th}$, and 84$^{th}$ percentiles. The two-dimensional distributions illustrate covariances between parameters; the contour lines approximate the 1$\sigma$ and 2$\sigma$ levels of the distributions.}
\figsetgrpend

\figsetgrpstart
\figsetgrpnum{15.5}
\figsetgrptitle{Binary members of Hyades}
\figsetplot{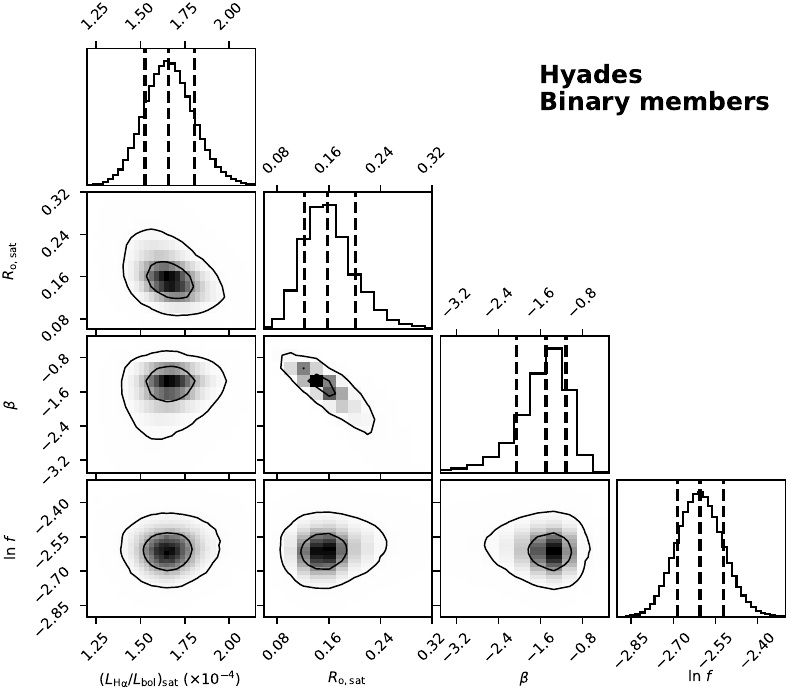}
\figsetgrpnote{Marginalized posterior probability distributions from the MCMC analysis of our \Ro--\LLH\ model using \texttt{emcee} for binary members of Hyades. The parameter values of the a posteriori model are the peaks of the one-dimensional distributions; the vertical dashed lines approximate the median and 16$^{th}$, 50$^{th}$, and 84$^{th}$ percentiles. The two-dimensional distributions illustrate covariances between parameters; the contour lines approximate the 1$\sigma$ and 2$\sigma$ levels of the distributions.}
\figsetgrpend

\figsetgrpstart
\figsetgrpnum{15.6}
\figsetgrptitle{Binary members of both clusters}
\figsetplot{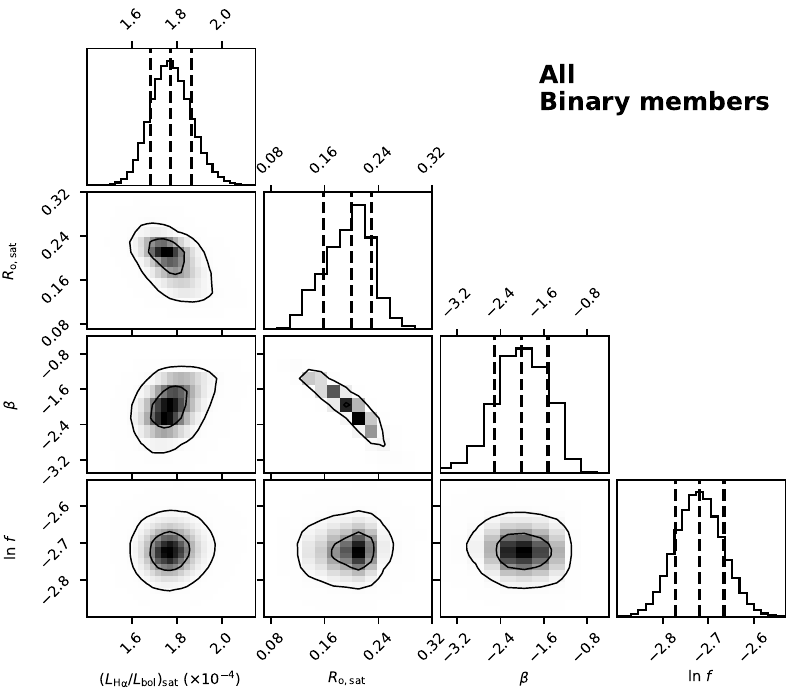}
\figsetgrpnote{Marginalized posterior probability distributions from the MCMC analysis of our \Ro--\LLH\ model using \texttt{emcee} for binary members of Praesepe and Hyades together. The parameter values of the a posteriori model are the peaks of the one-dimensional distributions; the vertical dashed lines approximate the median and 16$^{th}$, 50$^{th}$, and 84$^{th}$ percentiles. The two-dimensional distributions illustrate covariances between parameters; the contour lines approximate the 1$\sigma$ and 2$\sigma$ levels of the distributions.}
\figsetgrpend

\figsetend

\clearpage

\section{Re-Analysis of the \Ro--\LLH\ Relation Without Applying the Quiescent \halpha\ Absorption Correction}\label{app:noquiescent}

In this Appendix, we include the results we obtained after re-analyzing the \Ro--\LLH\ relation in \autoref{sec_rossbyresults} without correcting our \halpha\ EW measurements for the quiescent \halpha\ absorption present in these stars (see \autoref{sec_relativeEW}). This allows us to compare our findings to those of previous studies that did not include this correction, such as \citetalias{Douglas2014} and \citet{Nunez2017}.

In \citetalias{Douglas2014}, we found $\beta = -0.73^{+0.16}_{-0.12}$ for the combined sample of Praesepe and Hyades single members. Our new result in this Appendix, $\beta = -1.27^{+0.15}_{-0.17}$, is steeper, but within 2$\sigma$. In that previous study, we also noted a sharp decrease in \LLH\ over a small range in \Ro\ for stars with \Ro\ $\gtrsim$ 0.45. Whereas this claim was  speculative given the small sample size in \citetalias{Douglas2014}, our current expanded sample allows us to more confidently confirm it.

Our new \Ro$_\mathrm{,sat} = 0.14$$\pm$$0.01$ agrees within  1$\sigma$ with that found in \citetalias{Douglas2014}, \Ro$_\mathrm{,sat} = 0.11^{+0.02}_{-0.03}$. Notably, however, our new (\LLH)$_\mathrm{sat}$ = (1.73$\pm$0.06)$\times 10^{-4}$ disagrees with our previous result, (\LLH)$_\mathrm{sat}$ = (1.26$\pm$0.04)$\times 10^{-4}$, at the 5$\sigma$ level. This discrepancy is caused by the updated $\chi$ values that we used to calculate \LLH\ in this work. As shown in \autoref{fig_chiscomp}, our updated $\chi$ values are $\approx$1.3$\times$ larger than those used in \citetalias{Douglas2014}.

In \citet{Nunez2017}, we found $\beta = -0.51$$\pm$$0.02$ for a sample of single cluster members in the $\approx$500-Myr-old open cluster M37. Our $\beta$ in this Appendix for the combined sample of Praesepe and Hyades disagrees with the M37 result at the 5$\sigma$ level. We point out two issues that may be driving this large difference. First, the sample of \halpha\ EW measurements in the M37 study was partly contaminated by \halpha\ emission from a foreground  nebula. In that work, an attempt was made to mitigate the impact of the \halpha\ nebular emission by excluding stars for which [N \textsc{ii}] emission $\leq-3$ \AA. However, this still left stars with mildly contaminated \halpha\ EW values in the sample, leading to artificially higher \LLH\ values for those stars and potentially a shallower value for the best-fit $\beta$.

Second, the largest \Ro\ value for a M37 member in that study is $\approx$0.4, while our study shows the sharpest \LLH\ decline at \Ro\ $\gtrsim$ 0.4. In addition, most of the M37 stars with the lowest \LLH\ and largest \Ro\ values in the unsaturated regime have \LLH\ uncertainties larger than those in our sample. The MCMC algorithm that calculates $\beta$ incorporates the uncertainty associated with each measurement by assigning weights to individual data points. Therefore, these larger \LLH\ uncertainties in the M37 sample probably resulted in a shallower $\beta$.

Our new \Ro$_\mathrm{,sat}$ for the combined sample of Praesepe and Hyades stars is larger than the M37 \Ro$_\mathrm{,sat} = 0.03\pm0.01$ by a factor of almost 5. Also, our new (\LLH)$_\mathrm{sat}$ disagrees with the M37 (\LLH)$_\mathrm{sat}$ = (1.27$\pm$0.01)$\times 10^{-4}$ at the $>$5$\sigma$ level. Although the latter discrepancy is mostly explained by the aforementioned differences in $\chi$ values between the two studies, we found no evident explanation for the discrepancy in \Ro$_\mathrm{,sat}$. Outdated M37 stellar parameters, including cluster membership, may be partly driving the large differences in the characterization of the \Ro--\LLH\ relation between our study and that of M37.

\begin{figure*}[t]
\centerline{\includegraphics{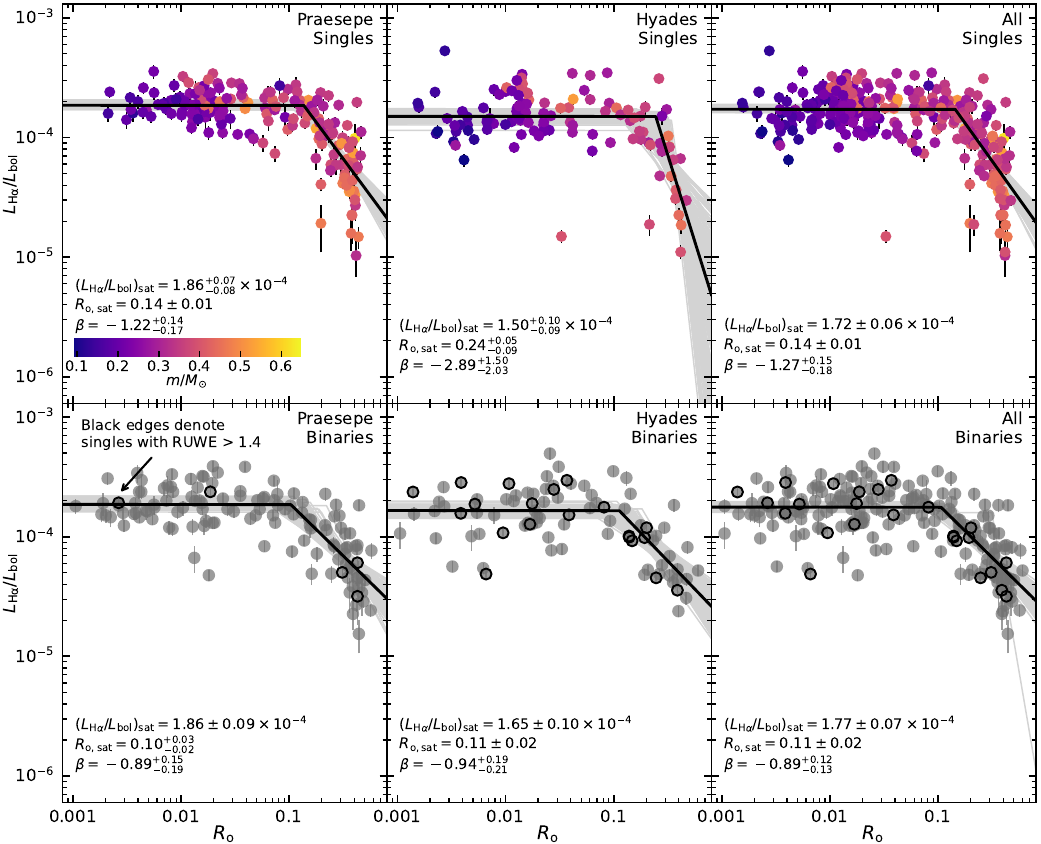}}
\caption{Same as \autoref{fig_rossbyHa}, but with \LLH\ calculated using our measured EW values instead of relative EW values from \autoref{sec_relativeEW}. The latter account for the \halpha\ quiescent photospheric absorption present in these stars.}
\label{fig_rossbyHanoquiesc}
\end{figure*}

\end{document}

%% file: tbl_measurements.tex
\begin{deluxetable}{@{}ll}
\tabletypesize{\scriptsize} 

\tablecaption{Overview of Columns in the Praesepe and Hyades
Membership Catalog \label{tbl_measurements}}

\tablehead{
\colhead{Column} & \colhead{Description\tablenotemark{a}} %\\[-0.1 in]
}

\startdata
1      & Name \\
2      & 2MASS designation \\
3      & Gaia DR3 designation \\
4      & Cluster to which the star belongs \\
5      & Binary flag: (0) no evidence for binarity; (1) candidate \\
       & binary; (2) confirmed binary \\
6, 7   & X-ray energy flux \fx\ (0.1--2.4 keV) and 1$\sigma$ uncertainty \\
8      & Rotation period \prot \\
9      & Source of \prot\tablenotemark{b} \\
10, 11 & Measured \halpha\ equivalent width EW and 1$\sigma$ uncertainty \\
12     & Number of spectra measured to obtain \halpha\ EW \\
13     & Relative \halpha\ EW \\
14, 15 & Effective temperature \teff\ and 1$\sigma$ uncertainty \\
16, 17 & $\chi$ and 1$\sigma$ uncertainty \\
18, 19 & Stellar radius $R_\mathrm{\star}$ and 1$\sigma$ uncertainty \\
\enddata

\tablenotetext{a}{This Table includes columns from Table 2 of \citet{Nunez2022} that have been updated and new columns from this study.}
\tablenotetext{b}{Possible values: ``TESS'' (period measurement from TESS data); ``ZTF;'' ``T\&Z'' (from both TESS and ZTF); ``Legacy'' (from \citealt{Douglas2019} or \citealt{Rampalli2022})}
\tablecomments{(This table is available in its entirety in machine-readable form.)}

\vspace{-0.2in}

\end{deluxetable}

%% file: tbl_MDMlog.tex
\setlength{\tabcolsep}{2.5pt}
\begin{deluxetable}{lccc}
\tablewidth{0pt}
\tabletypesize{\normalsize}
\tablecaption{Spectra of Praesepe and Hyades Stars Obtained Since \cite{Douglas2014}  \label{mdmstats}}
\tablehead{
\colhead{} & \colhead{} & \multicolumn{2}{c}{\# of Spectra} \\[-0.08 in]
\colhead{Dates} & \colhead{Instrument} & \colhead{Hyades} & \colhead{Praesepe}
}
\startdata
2014 Nov 10--Nov 16 & ModSpec & 6 & \nodata \\
2015 Feb 20--Feb 24 & ModSpec & \nodata & 20 \\ 
2015 Nov 21--Nov 22 & Hectospec & \nodata & 164 \\
2015 Dec 14--Dec 21 & ModSpec & 17 & 44 \\
2016 Jan 29--Feb 03 & ModSpec & 26 & 40 \\
2016 Nov 30--Dec 09 & ModSpec & 39 & 26 \\ 
2017 Feb 15          & ModSpec & \nodata & 8 \\
2017 Dec 15          & OSMOS & \nodata & 17 \\
2018 Jan 11--Jan 28 & OSMOS & 5 & 276\\
2018 Feb 04--Feb 09 & OSMOS & \nodata & 71 \\
2019 Jan 07--Jan 11 & OSMOS & 12 & \nodata \\
2019 Feb 27--Mar 02 & OSMOS & 14 & \nodata \\
2019 Nov 23--Nov 26 & OSMOS & 25 & \nodata \\
2021 Sep 05--Sep 29 & OSMOS & 14 & \nodata \\
2021 Oct 28          & OSMOS & 7  & \nodata \\
2022 Sep 01--Sep 02 & OSMOS & 10 & \nodata \\
2022 Sep 28--Oct 05 & OSMOS & 11 & \nodata \\
2022 Nov 10--Nov 14 & OSMOS & 61 & \nodata \\
2023 Mar 28--Mar 31 & OSMOS & 3  & \nodata\\
\hline
 & Total & 250 & 666 
\enddata
\tablecomments{All dates are UT.}

\vspace{-0.1in}

\end{deluxetable}

%2010 Dec 02--Dec 06 & ModSpec & \nodata & 107 \\
%2011 Feb 08--Feb 10 & ModSpec & \nodata & 60 \\
%2011 Nov 30--Dec 05 & ModSpec & 67 & \nodata \\
%2012 Feb 19--Feb 21 & ModSpec & 15 & 44 \\
%2012 Nov 11--Nov 14 & ModSpec & 65 & 7 \\

%% file: tbl_Xobs.tex
\setlength{\tabcolsep}{3pt}
\begin{deluxetable*}{cccrlcr}[th]
\tablewidth{0pt}
\tabletypesize{\normalsize}
\tablecaption{New Chandra Observations of Hyades Stars \label{chandra}}
\tablehead{
\colhead{Obs.~ID} & \multicolumn{2}{c}{Nominal Aimpoint} & \colhead{Roll} & \colhead{Target(s)} & \colhead{Start} & \colhead{Duration\tablenotemark{a}} \\[-0.05 in]
\cline{2-3}\\[-0.2 in]
\colhead{} & \colhead{$\alpha_{\rm J2000}$} & \colhead{$\delta_{\rm J2000}$} & \colhead{(\arcdeg)} & \colhead{(Gaia DR3 Desig.)} & \colhead{Date} & \colhead{(s)}
}
\startdata
27553 & 03:18:15.13 & +09:14.38.0 & 343.1 & \object[Gaia DR3 14143675198789504]{14143675198789504}\tablenotemark{b} & 2022-11 & 10085 \\
27566 & 03:13:03.29 & +32:53:55.2 & 210.1 & \object[Gaia DR3 125343573948444800]{125343573948444800}, & 2022-11 & 10083 \\
      &             &             &       & \object[Gaia DR3 125343608307015296]{125343608307015296} & & \\
27572 & 04:38:56.73 & +14:06:11.4 & 14.9  & \object[Gaia DR3 3309170875916905856]{3309170875916905856} & 2022-12 & 9902 \\
27573 & 04:33:41.90 & +19:00:38.0 & 28.5  & \object[Gaia DR3 3410453489022728576]{3410453489022728576} & 2022-12 & 9903 \\
27574 & 04:32:40.40 & +19:06:39.5 & 18.9  & \object[Gaia DR3 3410640887035452928]{3410640887035452928}, & 2022-12 & 9903 \\
      &             &             &       & \object[Gaia DR3 3410639993682264960]{3410639993682264960} & & \\
27612 & 04:48:50.99 & +15:56:57.9 & 299.8 & \object[Gaia DR3 3405127244241184256]{3405127244241184256} & 2022-12 & 10080 \\
27622 & 05:30:14.19 & +20:38:20.7 & 282.1 & \object[Gaia DR3 3402090466142958464]{3402090466142958464} & 2022-12 & 10083 \\
27623 & 06:03:26.87 & +24:02:26.8 & 265.5 & \object[Gaia DR3 3426209215771371648]{3426209215771371648} & 2022-12 & 20085 \\
28490 & 06:50:34.35 & $-$17:11:50.5 & 119.4 & \object[Gaia DR3 2946050323261707648]{2946050323261707648} & 2023-08 & 11082 \\
29061 & 04:16:13.11 & +18:53:04.2 & 91.2 & \object[Gaia DR3 47804394753757056]{47804394753757056,} & 2023-11 & 9945 \\
      &             &             &       & \object[Gaia DR3 47803952373768960]{47803952373768960} & & \\
29076 & 04:28:40.63 & +26:13:04.4 & 124.6 & \object[Gaia DR3 151222023217990016]{151222023217990016}\tablenotemark{b} & 2023-11 & 10086 \\
29088 & 03:50:03.26 & +22:35:43.0 & 268.0 & \object[Gaia DR3 64115585330656000]{64115585330656000} & 2023-12 & 10941 \\
29111 & 04:47:09.56 & +24:01:22.4 & 257.1 & \object[Gaia DR3 146989143968434688]{146989143968434688}\tablenotemark{b} & 2023-12 & 10086 \\
29112 & 04:47:41.72 & +26:09:11.2 & 244.1 & \object[Gaia DR3 154257259425702144]{154257259425702144}\tablenotemark{b} & 2023-12 & 10086 \\
\enddata
\tablenotetext{a}{Exposure time before any filtering is applied.}
\tablenotetext{b}{Undetected in observation.}

\vspace{-0.2in}

\end{deluxetable*}

%% file: tbl_newX.tex
\begin{deluxetable}{@{}ll}
\tabletypesize{\scriptsize} 

\tablecaption{Overview of Columns in the Addendum to the Praesepe and Hyades X-ray Source Catalog \label{tbl_newX}}

\tablehead{
\colhead{Column} & \colhead{Description\tablenotemark{a}} %\\[-0.1 in]
}

\startdata
1      & External catalog source ID \\
2      & Provenance of X-ray information\tablenotemark{b}\\
3      & IAU Name \\
4      & Observation ID \\
5      & Instrument \\
6, 7   & R.A., Decl. for epoch J2000\\
8      & X-ray positional uncertainty \\
9      & Off-axis angle $\theta$\\
10     & Detection likelihood $L$\tablenotemark{c} \\
11     & Net counts in the broad band \\
12, 13 & Net count rate and 1$\sigma$ uncertainty in broad band \\
14, 15 & Net count rate and 1$\sigma$ uncertainty in soft band \\
16, 17 & Net count rate and 1$\sigma$ uncertainty in hard band \\
18$-$20  & Definition of broad, soft, and hard bands \\
21     & Hardness ratio: (hard band $-$ soft band) / \\
       & (hard band + soft band) \\
22     & Exposure time \\
23     & Variability flag: (0) no evidence for variability; \\
       & (1) possibly variable; (2) definitely variable \\
24, 25 & Unabsorbed energy flux and 1$\sigma$ uncertainty in the \\
       & 0.1--2.4 keV band \\
26     & Source of energy flux: (ECF) from applying ECF; \\
       & (SpecFit) from spectral fitting \\
27     & X-ray flare removed? \\
28     & Quality Flag\tablenotemark{d} \\
29     & Name of the optical counterpart \\
30     & Separation between X-ray source and optical counterpart \\
\enddata

\tablenotetext{a}{This Table has the same columns and formats as those in Table 3 of \citet{Nunez2022}.}
\tablenotetext{b}{4XMM; CSC; CIAO: Reduction of Chandra observation with CIAO.}
\tablenotetext{c}{For CIAO sources, it is the source significance; for all others, it is the maximum likelihood.}
\tablenotetext{d}{m: likely mismatch to optical counterpart; x: likely extended source.}
\tablecomments{(This table is available in its entirety in machine-readable form.)}
\vspace{-0.4in}

\end{deluxetable}

%% file: tbl_chi.tex
\setlength{\tabcolsep}{3.5pt}
\begin{deluxetable}{cc@{\hskip 20pt}cc@{\hskip 20pt}cc}
\tablewidth{0pt}
\tabletypesize{\normalsize}
\tablecaption{\teff\ and $\chi$ Values from PHOENIX Model Spectra \label{tbl_chi}}
\tablehead{
\colhead{\teff} & \colhead{\hspace{-17pt}$\chi$} & \colhead{\teff} & \colhead{\hspace{-17pt}$\chi$} & \colhead{\teff} & \colhead{$\chi$} \\[-0.06 in]
\colhead{(K)} & \colhead{\hspace{-15pt}($\times 10^{-5}$)} & \colhead{(K)} & \colhead{\hspace{-15pt}($\times 10^{-5}$)} & \colhead{(K)} & \colhead{($\times 10^{-5}$)}
}
\startdata
6500 & 8.695 & 5000 & 9.581 & 3500 & 5.019 \\
6400 & 8.797 & 4900 & 9.409 & 3400 & 4.477 \\
6300 & 8.907 & 4800 & 9.279 & 3300 & 3.913 \\
6200 & 9.038 & 4700 & 9.195 & 3200 & 3.328 \\
6100 & 9.188 & 4600 & 9.127 & 3100 & 2.714 \\
6000 & 9.299 & 4500 & 9.071 & 3000 & 2.181 \\
5900 & 9.426 & 4400 & 8.658 & 2900 & 1.702 \\
5800 & 9.510 & 4300 & 8.197 & 2800 & 1.252 \\
5700 & 9.667 & 4200 & 7.632 & 2700 & 0.886 \\
5600 & 9.777 & 4100 & 7.144 & 2600 & 0.618 \\
5500 & 9.351 & 4000 & 6.825 & 2500 & 0.473 \\
5400 & 8.961 & 3900 & 6.523 & 2400 & 0.603 \\
5300 & 8.597 & 3800 & 6.201 & 2300 & 0.564 \\
5200 & 9.160 & 3700 & 5.858 &  &  \\
5100 & 9.622 & 3600 & 5.494 &  & 
\enddata
\tablecomments{The methodology used to calculate $\chi$ is described in the appendix of \citetalias{Douglas2014}.}

\vspace{-0.2in}

\end{deluxetable}

%% file: tbl_Rossbies.tex
\begin{deluxetable*}{@{}rrcccrccccc@{}}
\tabletypesize{\small}

\tablecaption{Rotation--Activity Relation Fitting Results\label{tbl_rossbies}}

\tablehead{
 & \multicolumn{4}{c}{\Ro--\LLH} & \multicolumn{6}{c}{\Ro--\LLX}\\
\cmidrule(lr){2-5}\cmidrule(lr){6-11} \\[-0.6cm]
\colhead{Sample} & \colhead{$N_\star$} & \colhead{(\LLH)$_\mathrm{sat}$} & \colhead{$R_\mathrm{o,sat}$} & \colhead{$\beta$} & \colhead{$N_\star$} & \colhead{$\beta_\mathrm{sup}$} & \colhead{$R_\mathrm{o,sup}$} & \colhead{(\LLX)$_\mathrm{sat}$} & \colhead{$R_\mathrm{o,sat}$} & \colhead{$\beta$}\\[-0.06 in]
 & & \colhead{($10^{-4}$)} & & & & & & \colhead{($10^{-3}$)} & & \\[-0.2 in]
}
\startdata
\multicolumn{11}{c}{\it Single stars} \\[0.02 in]
Praesepe & 196 & 1.76$\pm$0.09 & 0.28$^{+0.02}_{-0.03}$ & $-5.19^{+1.32}_{-0.94}$ & 124 & 0.70$^{+0.60}_{-0.32}$ & 0.011$\pm$0.005 & 1.14$\pm$0.12 & 0.19$\pm$0.02 & $-3.48^{+0.34}_{-0.39}$\\
Hyades   & 116 & 1.53$\pm$0.08 & 0.31$^{+0.01}_{-0.02}$ & $-7.07^{+1.40}_{-1.57}$ & 162 & 0.54$^{+0.19}_{-0.15}$ & 0.014$^{+0.004}_{-0.005}$ & 1.15$^{+0.13}_{-0.12}$ & 0.17$\pm$0.02 & $-3.04^{+0.27}_{-0.28}$ \\
All      & 312 & 1.65$\pm$0.06 & 0.29$\pm$0.01 & $-5.85^{+0.81}_{-0.80}$ & 286 & 0.53$^{+0.16}_{-0.12}$ & 0.015$^{+0.003}_{-0.005}$ & 1.17$\pm$0.09 & 0.17$\pm$0.01 & $-3.18^{+0.20}_{-0.21}$ \\[0.05 in]
\multicolumn{11}{c}{\it Binaries \& stars with RUWE $>$ 1.4} \\[0.02 in]
Praesepe & 134 & 1.84$^{+0.11}_{-0.12}$ & 0.24$\pm$0.03 & $-2.77^{+0.51}_{-0.70}$ & 112 & 0.13$^{+0.22}_{-0.11}$ & 0.009$^{+0.007}_{-0.004}$ & 1.17$\pm$0.15 & 0.12$^{+0.03}_{-0.02}$ & $-2.20^{+0.26}_{-0.38}$ \\
Hyades   &  92 & 1.63$\pm$0.14 & 0.16$^{+0.05}_{-0.04}$ & $-1.51^{+0.39}_{-0.59}$ & 138 & 0.08$^{+0.15}_{-0.07}$ & 0.009$^{+0.007}_{-0.005}$ & 1.02$^{+0.11}_{-0.10}$ & 0.14$\pm$0.02 & $-2.36^{+0.27}_{-0.31}$ \\
All      & 226 & 1.76$\pm$0.09 & 0.20$^{+0.03}_{-0.04}$ & $-2.05\pm$0.50 & 250 & 0.06$^{+0.11}_{-0.05}$ & 0.009$^{+0.007}_{-0.005}$ & 1.07$\pm$0.08 & 0.13$^{+0.02}_{-0.01}$ & $-2.26^{+0.19}_{-0.24}$ \\ [0.05 in]
\enddata

\vspace{-0.5in}

\end{deluxetable*}